\documentclass[
prd,
reprint,
nofootinbib,
superscriptaddress,
 amsmath,
 amssymb,
 aps,
floatfix,
showkeys
]{revtex4-1}

\makeatletter
\def\@fnsymbol#1{\ensuremath{\ifcase#1\or *\or \dagger\or \ddagger\or
   \mathsection\or \mathparagraph\or \|\or **\or \dagger\dagger
   \or \ddagger\ddagger \or *** \else\@ctrerr\fi}}
\makeatother

\usepackage[colorlinks=true,citecolor=blue,filecolor=blue,linkcolor=blue,urlcolor=blue]{hyperref}
\usepackage{graphicx}
\usepackage{tabularx}
\usepackage{dcolumn}
\usepackage{bm}
\usepackage{xcolor}
\usepackage{float}
\usepackage{adjustbox}
\usepackage{lineno}
\usepackage{array}
\usepackage{siunitx}
\usepackage{enumerate}
\usepackage{multirow}
\usepackage{url}
\usepackage{upgreek}
\usepackage{physics}
\newcommand{\lal}{\affiliation{Universit\'{e} Paris-Saclay, CNRS/IN2P3, IJCLab, 91405 Orsay, France}}
\newcommand{\mainz}{\affiliation{Institut f\"ur Physik \& Exzellenzcluster PRISMA, Johannes Gutenberg-Universit\"at Mainz, 55099 Mainz, Germany}}
\newcommand{\torino}{\affiliation{INAF-Astrophysical Observatory of Torino, Department of Physics, University  of  Torino and  INFN-Torino,  10125  Torino,  Italy}}
\newcommand{\columbia}{\affiliation{Physics Department, Columbia University, New York, NY 10027, USA}}
\newcommand{\ucla}{\affiliation{Physics \& Astronomy Department, University of California, Los Angeles, CA 90095, USA}}
\newcommand{\zurich}{\affiliation{Physik-Institut, University of Z\"urich, 8057  Z\"urich, Switzerland}}
\newcommand{\nyuad}{\affiliation{New York University Abu Dhabi, Abu Dhabi, United Arab Emirates}}
\newcommand{\subatech}{\affiliation{SUBATECH, IMT Atlantique, CNRS/IN2P3, Universit\'e de Nantes, Nantes 44307, France}}

\newcommand{\wis}{\affiliation{Department of Particle Physics and Astrophysics, Weizmann Institute of Science, Rehovot 7610001, Israel}}
\newcommand{\nagoya}{\affiliation{Kobayashi-Maskawa Institute for the Origin of Particles and the Universe, and Institute for Space-Earth Environmental Research, Nagoya University, Furo-cho, Chikusa-ku, Nagoya, Aichi 464-8602, Japan}}
\newcommand{\coimbra}{\affiliation{LIBPhys, Department of Physics, University of Coimbra, 3004-516 Coimbra, Portugal}}
\newcommand{\rpi}{\affiliation{Department of Physics, Applied Physics and Astronomy, Rensselaer Polytechnic Institute, Troy, NY 12180, USA}}
\newcommand{\kobe}{\affiliation{Department of Physics, Kobe University, Kobe, Hyogo 657-8501, Japan}}

\newcommand{\purdue}{\affiliation{Department of Physics and Astronomy, Purdue University, West Lafayette, IN 47907, USA}}
\newcommand{\napels}{\affiliation{Department of Physics ``Ettore Pancini'', University of Napoli and INFN-Napoli, 80126 Napoli, Italy}}

\newcommand{\tokyo}{\affiliation{Kamioka Observatory, Institute for Cosmic Ray Research, and Kavli Institute for the Physics and Mathematics of the Universe (WPI), the University of Tokyo, Higashi-Mozumi, Kamioka, Hida, Gifu 506-1205, Japan}}
\newcommand{\laquila}{\affiliation{Department of Physics and Chemistry, University of L'Aquila, 67100 L'Aquila, Italy}}
\newcommand{\chicago}{\affiliation{Department of Physics \& Kavli Institute for Cosmological Physics, University of Chicago, Chicago, IL 60637, USA}}
\newcommand{\paris}{\affiliation{LPNHE, Sorbonne Universit\'{e}, Universit\'{e} de Paris, CNRS/IN2P3, Paris, France}}
\newcommand{\bologna}{\affiliation{Department of Physics and Astronomy, University of Bologna and INFN-Bologna, 40126 Bologna, Italy}}
\newcommand{\munster}{\affiliation{Institut f\"ur Kernphysik, Westf\"alische Wilhelms-Universit\"at M\"unster, 48149 M\"unster, Germany}}
\newcommand{\nikhef}{\affiliation{Nikhef and the University of Amsterdam, Science Park, 1098XG Amsterdam, Netherlands}}

\newcommand{\stockholm}{\affiliation{Oskar Klein Centre, Department of Physics, Stockholm University, AlbaNova, Stockholm SE-10691, Sweden}}
\newcommand{\freiburg}{\affiliation{Physikalisches Institut, Universit\"at Freiburg, 79104 Freiburg, Germany}}
\newcommand{\rice}{\affiliation{Department of Physics and Astronomy, Rice University, Houston, TX 77005, USA}}
\newcommand{\lngs}{\affiliation{INFN-Laboratori Nazionali del Gran Sasso and Gran Sasso Science Institute, 67100 L'Aquila, Italy}}
\newcommand{\heidelberg}{\affiliation{Max-Planck-Institut f\"ur Kernphysik, 69117 Heidelberg, Germany}}
\newcommand{\ucsd}{\affiliation{Department of Physics, University of California San Diego, La Jolla, CA 92093, USA}}

\newcommand{\cea}{\affiliation{CEA, LIST, Laboratoire National Henri Becquerel, CEA-Saclay 91191 Gif-sur-Yvette Cedex, France}}

\begin{document}

\title{Excess Electronic Recoil Events in XENON1T}

\author{E.~Aprile}\columbia
\author{J.~Aalbers}\stockholm
\author{F.~Agostini}\bologna
\author{M.~Alfonsi}\mainz
\author{L.~Althueser}\munster
\author{F.~D.~Amaro}\coimbra
\author{V.~C.~Antochi}\stockholm
\author{E.~Angelino}\torino
\author{J.~R.~Angevaare}\nikhef
\author{F.~Arneodo}\nyuad
% ...
\collaboration{XENON Collaboration}
\email[]{xenon@lngs.infn.it}
\noaffiliation
\author{\vspace{-0.3cm}X.~Mougeot}\cea

\date{\today}

\begin{abstract}

We report results from searches for new physics with low-energy electronic recoil data recorded with the XENON1T detector. With an exposure of 0.65 tonne-years and an unprecedentedly low background rate of \mbox{$76 \pm 2_{\,\mathrm{stat}}$\,events/(tonne $\times$ year $\times$ keV)} between 1--30\,keV, the data enables one of the most sensitive searches for solar axions, an enhanced neutrino magnetic moment using solar neutrinos, and bosonic dark matter. An excess over known backgrounds is observed at low energies and most prominent between 2--3\,keV. The solar axion model has a 3.4\,$\sigma$ significance, and a three-dimensional 90\% confidence surface is reported for axion couplings to electrons, photons, and nucleons. This surface is inscribed in the cuboid defined by $g_{\mathrm{ae}} < 3.8 \times 10^{-12}$, $g_{\mathrm{ae}}g_{\mathrm{an}}^\mathrm{eff} < 4.8 \times 10^{-18}$, and $g_{\mathrm{ae}}g_{\mathrm{a\upgamma}} < 7.7\times10^{-22}~\mathrm{GeV}^{-1}$, and excludes either $g_{\mathrm{ae}}=0$ or $g_{\mathrm{ae}}g_\mathrm{a\upgamma}=g_{\mathrm{ae}}g_{\mathrm{an}}^{\mathrm{eff}}=0$. The neutrino magnetic moment signal is similarly favored over background at 3.2\,$\sigma$ and a confidence interval of  $\mu_{\nu} \in (1.4,~2.9) \times 10^{-11}\,\mu_B$ (90\% C.L.) is reported. Both results are in strong tension with stellar constraints. The excess can also be explained by \mbox{$\upbeta$ decays} of tritium at 3.2\,$\sigma$ significance with a corresponding tritium concentration in xenon of $(6.2 \pm 2.0) \times 10^{-25}$\,mol/mol. Such a trace amount can neither be confirmed nor excluded with current knowledge of its production and reduction mechanisms. The significances of the solar axion and neutrino magnetic moment hypotheses are decreased to 2.0\,$\sigma$ and 0.9\,$\sigma$, respectively, if an unconstrained tritium component is included in the fitting.  With respect to bosonic dark matter, the excess favors a monoenergetic peak at ($2.3 \pm 0.2$)\,keV (68\% C.L.) with a 3.0\,$\sigma$ global (4.0\,$\sigma$ local) significance over background. This analysis sets the most restrictive direct constraints to date on pseudoscalar and vector bosonic dark matter for most masses between 1~and~210\,keV/c$^2$. We also consider the possibility that $^{37}$Ar may be present in the detector, yielding a $2.82$\,keV peak from electron capture. Contrary to tritium, the $^{37}$Ar concentration can be tightly constrained and is found to be negligible.

\begin{description}
\item[PACS numbers]
\item[Keywords]
\end{description}
\end{abstract}

\pacs{
    95.35.+d, 
    14.80.Ly, 
    29.40.-n,  
    95.55.Vj
}

\keywords{Dark Matter, Direct Detection, Xenon}

\maketitle


\section{Introduction}
A preponderance of astrophysical and cosmological evidence suggests that most of the matter content in the Universe is made up of a rarely interacting, non-luminous component called dark matter~\cite{BertoneHooper2005}. Although several hypothetical dark matter particle candidates have been proposed with an assortment of couplings, masses, and detection signatures, dark matter has thus far eluded direct detection. The XENON1T experiment~\cite{xe1t_instrument}, employing a liquid-xenon time projection chamber (LXe TPC), was primarily designed to detect Weakly Interacting Massive Particle (WIMP) dark matter. Due to its unprecedentedly low background rate, large target mass, and low energy threshold, XENON1T is also sensitive to interactions from alternative dark matter candidates and to other physics beyond the Standard Model (SM). Here we report on searches for (1) axions produced in the Sun, (2) an enhancement of the neutrino magnetic moment using solar neutrinos, and (3) pseudoscalar and vector bosonic dark matter, including axion-like particles and dark photons.

The XENON1T experiment operated underground at the INFN Laboratori Nazionali del Gran Sasso (LNGS) from 2016--2018, utilizing a \mbox{dual-phase} LXe TPC with a \mbox{2.0-tonne} active target to search for rare processes. A particle interaction within the detector produces both prompt scintillation (S1) and delayed electroluminesence\,(S2) signals. These light signals are detected by arrays of photomultiplier tubes (PMTs) on the top and bottom of the active volume, and are used to determine the deposited energy and interaction position of an event. The latter allows for removing background events near the edges of the target volume (e.g., from radioactivity in detector materials) through fiducialization. The S2/S1 ratio is used to distinguish electronic recoils (ERs), produced by, e.g., gamma rays ($\upgamma$s) or beta electrons ($\upbeta$s), from nuclear recoils (NRs), produced by, e.g.,~neutrons or WIMPs, allowing for a degree of particle identification. The ability to determine scatter multiplicity enables further reduction of backgrounds, as signals are expected to have only single energy deposition.

In this paper, we report on searches for ER signals with data acquired from February 2017 to February 2018, a time period referred to as Science Run 1 (SR1)~\cite{xe1t_sr0_sr1}. As the vast majority of background comes from ER events, we search for excesses above a known background level. The analysis is carried out in the space of reconstructed energy, which exploits the anti-correlation of S1 and S2 signals by combining them into a single energy scale~\cite{1t_highe}, thus reducing the statistical fluctuations from \mbox{electron-ion} recombination~\cite{Aprile:2007qd}. Both S1 and S2 signals are corrected to disentangle position-dependent effects, such as light collection efficiency (LCE) and electron attachment to electronegative impurities. After correcting to the mean LCE across the TPC, S1 is reconstructed using signals from all PMTs (cS1). For the S2 reconstruction, only the bottom PMT array is used (cS2$_\mathrm{b}$) because it features a more homogeneous light collection~\cite{xe1t_sr0_sr1}. The full energy region of interest (ROI) is (1, 210) keV, which is primarily motivated by the search for bosonic dark matter and discussed further in Sec.~\ref{subsec:dataselection}. 

The paper is organized as follows. In Sec.~\ref{sec:signalmodel} we present the theoretical background and signal modeling of the \mbox{beyond-the-SM} channels considered in this search. We describe the data analysis in Sec.~\ref{sec:analysis}, including the data selection, background model, and statistical framework. In Sec.~\ref{sec:results}, upon observation of a \mbox{low-energy} excess in the data, we present a hypothesis of a new background component, tritium, which may be observable for the first time in a xenon detector due to our unprecedented low background. We then report the results of searches for solar axions, an anomalous neutrino magnetic moment, and bosonic dark matter. We end with further discussion of our findings and a summary of this work in Secs.~\ref{sec:discussion} and \ref{sec:summary}, respectively. The presence of the excess motivated further scrutiny of the modeling of dominant backgrounds, the details of which we present in the Appendix.

\section{Signal Models}
\label{sec:signalmodel}
This section describes the physics channels we search for in this work. In Sec.~\ref{subsec:solaraxion}, we motivate the search of solar axions, presenting their production mechanisms in the Sun and the detection mechanism in LXe TPCs, and summarize two benchmark axion models. In Sec.~\ref{subsec:nmm}, we introduce the search for an anomalous neutrino magnetic moment, which would enhance the neutrino-electron elastic scattering cross section at low energies. In Sec.~\ref{subsec:bdm}, we discuss the signals induced by bosonic dark matter including pseudoscalar and vector bosons, examples of which are \mbox{axion-like} particles and dark photons, respectively. Expected energy spectra of these signals in the XENON1T detector are summarized at the end of this section.

For all signal models presented below, the theoretical energy spectra in a LXe TPC were converted to the space of reconstructed energy by accounting for detector efficiency and resolution, summarized in Fig.~\ref{fig:signals}.  The efficiency is shown in Fig.~\ref{fig:efficiency} and discussed in Sec.~\ref{subsec:dataselection}. For the energy resolution, the theoretical spectra were smeared using a Gaussian distribution with \mbox{energy-dependent} width, which was determined using an empirical fit of \mbox{mono-energetic} peaks as described in~\cite{xe1t_instrument,1t_highe}. The energy resolution $\sigma$ is given by
\begin{equation}
    \sigma(E) = a\cdot\sqrt{E} + b\cdot E,
    \label{eq:energy_resolution}
\end{equation}
with \mbox{$a$ = ($0.310 \pm 0.004$)\,$\sqrt{\mathrm{keV}}$ and $b$ = 0.0037 $\pm$ 0.0003}.

When building a signal model, the resolution is first applied to the deposited, ``true" energy spectrum, and then the smeared distribution is corrected according to the predicted loss due to efficiency. This implies that, near threshold, the mean reconstructed energy is higher than the true energy, as the reduced efficiency at lower energies shifts the mean of the observed distribution upwards. This type of reconstruction bias is fully accounted for in this analysis.

\subsection{Solar Axions}\label{subsec:solaraxion}
As a solution to the strong CP problem in quantum chromodynamics (QCD), Peccei and Quinn postulated a mechanism that naturally gives rise to a Nambu-Goldstone boson, the \mbox{so-called} axion~\cite{PecceiQuinn_1977, Weinberg, Wilczek}. In addition to solving the strong CP problem, QCD axions are also well-motivated dark matter candidates, with cosmological and astrophysical bounds requiring their mass to be small \mbox{(typically $\ll$\,keV})~\cite{PRESKILL1983127,ABBOTT1983133,Dine:1982ah,Cadamuro,Raffelt_bounds}. On account of this mass constraint, \emph{dark matter} axions produced in the early Universe cannot be observed in XENON1T. However, \emph{solar} axions would emerge with---and in turn deposit---energies in the keV range~\cite{Redondo:2013wwa, Shigetaka1995, vanBibber_1989}, the precise energies to which XENON1T was designed to be most sensitive. An observation of solar axions would be evidence of \mbox{beyond-the-SM} physics, but would not by itself be sufficient to draw conclusions about axionic dark matter.

We consider three production mechanisms that contribute to the total solar axion flux: (1) Atomic recombination and deexcitation, Bremsstrahlung, and Compton\,(ABC) interactions \cite{Redondo:2013wwa, Dimopoulos1986kc}, (2) a mono-energetic 14.4~keV M1 nuclear transition of $^{57}$Fe \cite{Shigetaka1995}, and (3) the Primakoff conversion of photons to axions in the Sun~\cite{Primakoff, Kolb_1978}. The ABC flux scales with the axion-electron coupling $g_{\mathrm{ae}}$ as
\begin{equation}
   \Phi_\mathrm{a}^{\mathrm{ABC}} \propto g_{\mathrm{ae}}^2
   \label{eq:abc_flux}
\end{equation}
and was taken from \cite{Redondo:2013wwa}. The $^{57}$Fe flux scales with an effective axion-nucleon coupling \mbox{$g_\mathrm{an}^\mathrm{eff}=-1.19g_\mathrm{an}^0 + g_\mathrm{an}^3$} and is given by \cite{fe57_cuore, CAST2009}
\begin{equation}
    \Phi_\mathrm{a}^{^{57}\mathrm{Fe}} = \left(\frac{k_\mathrm{a}}{k_\upgamma}\right)^3 \times 4.56 \times 10^{23} (g_\mathrm{an}^\mathrm{eff})^2~~\mathrm{cm}^{-2}\mathrm{s}^{-1},
\end{equation}
where $g_\mathrm{an}^{0/3}$ are the isoscalar/isovector coupling constants and $k_\mathrm{a}$ and $k_\gamma$ are the momenta of the produced axion and photon, respectively. The Primakoff flux scales with the axion-photon coupling $g_\mathrm{a\upgamma}$ and is given by~\cite{axiontheory}
\begin{align}
\begin{split}
    \frac{d\Phi_\mathrm{a}^{\mathrm{Prim}}}{dE_\mathrm{a}} = &\left( \frac{g_\mathrm{a\upgamma}}{\mathrm{GeV}^{-1}} \right)^2 \left( \frac{E_\mathrm{a}}{\mathrm{keV}} \right)^{2.481}e^{-E_\mathrm{a}/(1.205~\mathrm{keV})}\\
    &\times 6 \times 10^{30}~~\mathrm{cm}^{-2}\mathrm{s}^{-1}\mathrm{keV}^{-1},
\end{split}
\label{eq:primakoff_flux}
\end{align}
where $E_\mathrm{a}$ is the energy of the axion.
All three flux components could be detected in XENON1T via the axioelectric effect -- the axion analog to the photoelectric effect -- which has a cross section that scales with axion-electron coupling $g_{\mathrm{ae}}$ and is given by \cite{Dimopoulos:1986mi, Dimopoulos:1985tm, Pospelov2008a, fe57_cuore}
\begin{equation}
    \sigma_\mathrm{ae} = \sigma_\mathrm{pe} \frac{g_{\mathrm{ae}}^2}{\beta} \frac{3E_\mathrm{a}^2}{16\pi\alpha m_\mathrm{e}^2}\left(1 - \frac{\beta^{2/3}}{3} \right),
    \label{eq:pseudo_xsec}
\end{equation}
where $\beta$ and $E_\mathrm{a}$ are the velocity and energy of the axion, respectively, $\alpha$ is the fine structure constant, and $m_e$ is the mass of the electron. The energy-dependent photoelectric cross section, $\sigma_\mathrm{pe}$, was obtained from \cite{Veigele} and interpolated between points using the logarithms of both photon energies and cross sections. Combining the production and detection mechanisms, we are able to constrain the values of $\abs{g_{\mathrm{ae}}}$~(ABC), $\abs{g_{\mathrm{ae}}g_\mathrm{an}^\mathrm{eff}}$ ($^{57}$Fe), and $\abs{g_{\mathrm{ae}}g_\mathrm{a\upgamma}}$ (Primakoff)\footnote{We drop the absolute value notation for the remainder of this paper.}. We consider these three observables independently in the analysis, lest we implicitly assume any particular axion model. Still, it is important to note that these values are indeed related to each other and to the axion mass under different models. 

For QCD axions, the mass $m_\mathrm{a}$ is related to the decay constant $f_\mathrm{a}$ via
\begin{equation}
m_\mathrm{a} \simeq \frac{6 \times 10^6 \mathrm{~GeV}}{f_\mathrm{a}}~~{\mathrm{eV/c}^2},
\label{eq:mass_scale}
\end{equation}
and the axion couplings to matter are mostly \mbox{model-dependent}. We describe here two benchmark classes of QCD axion models: Dine-Fischler-Srednicki-Zhitnitsky\,(DFSZ)~\cite{dfs_1981, Zhitnitsky_1980}, in which axions couple to electrons at tree level, and Kim-Shifman-Vainshtein-Zhakharov\,(KSVZ)~\cite{kim_1979,svz_1980}, where couplings to leptons occur only at loop level. For this reason the ABC flux is dominant in DFSZ models, while the Primakoff flux is dominant in KSVZ models. Since the axioelectric cross section scales with the axion-electron coupling, XENON1T is in general more sensitive to DFSZ-type axions.

\begin{figure*}[!htbp]
    \centering
    \includegraphics[width=0.95\textwidth]{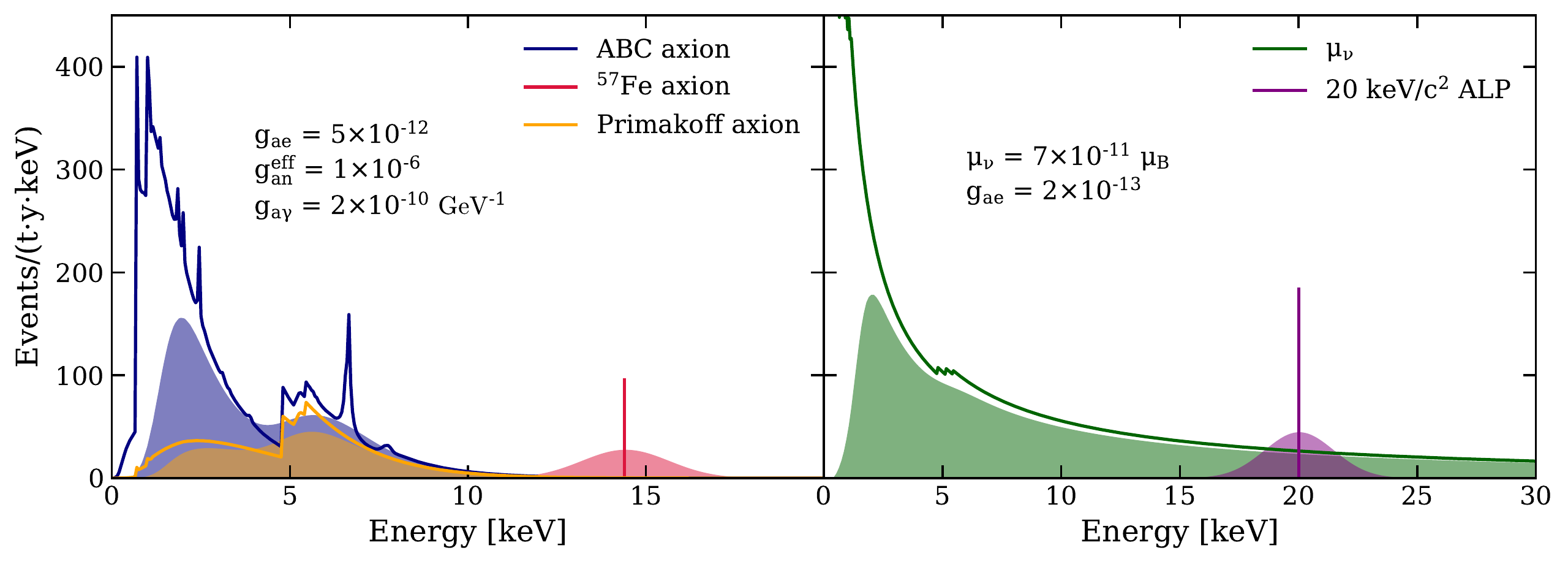}
    \caption{Left: Expected signal in energy space for ABC solar axions with a coupling $g_{\mathrm{ae}}=5\times 10^{-12}$ (blue), for solar axions produced from the de-excitation of $^{57}$Fe with coupling $g_\mathrm{an}^\mathrm{eff}=1\times 10^{-6}$ (red), and for solar axions produced from the Primakoff effect with coupling $g_\mathrm{a\upgamma}=2\times 10^{-10}$ (orange). Right: Signature of an enhanced neutrino magnetic moment with magnitude $7\times10^{-11}\,\mu_B$ (green) and a 20\,keV/c$^2$ ALP with coupling constant $g_{\mathrm{ae}}=2\times 10^{-13}$ (purple). Both the true deposited energy spectra in a xenon detector without efficiency loss (unshaded) and the expected observed spectra in XENON1T including the specific detector resolution and efficiency (shaded) are shown.}
    \label{fig:signals}
\end{figure*}

In DFSZ models the axion-electron coupling is given by 
\begin{equation}
    g_{\mathrm{ae}} = \frac{m_\mathrm{e}}{3f_\mathrm{a}}\cos^2 \beta_{\mathrm{DFSZ}},
\label{eq:gae}
\end{equation}
where 
\begin{equation}
\tan(\beta_\mathrm{DFSZ}) = \left(\frac{X_\mathrm{u}}{X_\mathrm{d}}\right)^{1/2}, 
\end{equation} 
and $X_\mathrm{u}$ and $X_\mathrm{d}$ are the Peccei-Quinn (PQ) charges of the up and down quarks, respectively~\cite{Srednicki_1985, pdg2019, fe57_cuore}. The couplings to quarks take on a similar expression with respect to $\beta_{\mathrm{DFSZ}}$. The axion-nucleon couplings $g_\mathrm{an}^0$ and $g_\mathrm{an}^3$ are functions of $X_\mathrm{u}$, $X_\mathrm{d}$, and $f_\mathrm{a}$, and can be found in \cite{Srednicki_1985, Kaplan_1985}. For a DFSZ axion, it follows that $g_{\mathrm{ae}}$ and $g_\mathrm{an}^\mathrm{eff}$ are both non-zero in general, as they are connected via $\beta_{\mathrm{DFSZ}}$ and $f_\mathrm{a}$. The axion-photon coupling does not depend on the PQ charges but is directly related to the axion decay constant (and thus the mass):
\begin{equation}
    g_\mathrm{a\upgamma} = \frac{\alpha}{2\pi f_\mathrm{a}} \left( \frac{E}{N} - \frac{2}{3}\: \frac{4 + z}{1 + z}\right),
\label{eq:ga_gamma}
\end{equation}
where $z = {m_\mathrm{u}} / {m_\mathrm{d}}$, $m_\mathrm{u/d}$ are the respective masses of the up/down quarks, and $E/N$ represent the \mbox{model-dependent} electromagnetic/color anomalies of the axial current associated with the axion field~\cite{di_Cortona_2016}. It is typically assumed that \mbox{$E/N = 8/3$} in DFSZ models. 

In KSVZ models, the PQ charges of the SM quarks vanish, and there is no $\beta_{\mathrm{DFSZ}}$-like parameter. The \mbox{axion-electron} coupling strength, induced by radiative corrections, depends on the axial current~\cite{Chang_Choi_1993, Srednicki_1985}:
\begin{align}
\begin{split}
    g_{\mathrm{ae}} = &\left( \frac{E}{N} \ln \frac{f_\mathrm{a}}{m_\mathrm{e}} - \frac{2}{3}\: \frac{4 + z + w}{1 + z + w} \ln \frac{\Lambda}{m_\mathrm{e}} \right)\\ &\times \frac{3\alpha^{2}Nm_{\mathrm{e}}}{4\pi^2 f_\mathrm{a}},
\label{eq:gae_ksvz}
\end{split}
\end{align}
where $w = {m_\mathrm{u}}/{m_\mathrm{s}}$, $m_\mathrm{s}$ is the mass of the strange quark; $\Lambda$ is the cutoff of the QCD confinement scale. The isoscalar/isovector axion-nucleon couplings $g_\mathrm{an}^{0/3}$ do not depend on the PQ charges and are also found in \cite{Srednicki_1985, Kaplan_1985}. The axion-photon coupling is given by Eq.~(\ref{eq:ga_gamma}). For KSVZ models a benchmark value of \mbox{$E/N=0$} is often used, but many values are possible~\cite{DiLuzio2017}. 

As mentioned above, no particular axion model is assumed in the analysis itself; the three flux components are considered completely independent of each other. Since, in principle, it is possible for all three components to be present at the same time, our solar axion model includes three unconstrained parameters for the different components. Were a signal observed, the results of the three-component analysis could then be used to constrain different axion models and possibly infer the axion mass. This approach also implies that the results hold generally for solar axion-like particles, which do not have strict relationships between the couplings, as described in Sec.~\ref{subsec:bdm}.

The expected spectra from solar axions with \mbox{$g_{\mathrm{ae}}=5\times10^{-12}$}, \mbox{$g_\mathrm{a\upgamma}=2\times10^{-10}~\mathrm{GeV}^{-1}$}, and \mbox{$g_\mathrm{an}^\mathrm{eff}=10^{-6}$} are shown in Fig.~\ref{fig:signals} (left) with before/after detector effects indicated by unshaded/shaded curves, respectively. The rate of the ABC component is proportional to ${g_{\mathrm{ae}}}^4$; the $^{57}$Fe component is proportional to $(g_{\mathrm{ae}}g_\mathrm{an}^\mathrm{eff})^2$; and the Primakoff component is proportional to $(g_{\mathrm{ae}}g_\mathrm{a\upgamma})^2$.

\subsection{Neutrino Magnetic Moment}\label{subsec:nmm}
In the SM, neutrinos are massless, and therefore without a magnetic dipole moment. However, the observation of neutrino oscillation tells us that neutrinos have mass and the SM must be extended, thus implying a magnetic moment of \mbox{$\mu_{\nu} \sim 10^{-20}\,\mu_B$}~\cite{Shrock,Kim_nmm_1976,Kim_nmm_1978, Bell_2006}, where $\mu_B$ is the Bohr magneton. Larger values of \mbox{$\mu_{\nu}$} have been considered theoretically and experimentally \cite{Bell_2005, Bell_2006, borexino_nmm}. Interestingly, in addition to providing evidence of \mbox{beyond-SM physics}, the observation of a $\mu_{\nu} \gtrsim 10^{-15}\,\mu_B$ would suggest that neutrinos are Majorana fermions~\cite{Bell_2006}. Currently the most stringent direct detection limit is \mbox{$\mu_{\nu} < 2.8 \times 10^{-11}\,\mu_B$} from Borexino~\cite{borexino_nmm}, and indirect constraints based on the cooling of globular cluster and white dwarfs are an order of magnitude stronger at $\sim 10^{-12}\,\mu_B$~\cite{nmm_rgb, pdg2019, redgiants_gae}.

An enhanced magnetic moment would increase the neutrino scattering cross-sections at low energies (on both electrons and nuclei), and thus could be observable by low-threshold detectors such as XENON1T. Here we only consider the enhancement to elastic scattering on electrons, given by \cite{nmm}
\begin{equation}
\label{eq:mu_xsec}
\frac{d\sigma_{\mu}}{dE_\mathrm{r}} = \mu_{\nu}^2 \alpha \left(\frac{1}{E_\mathrm{r}} - \frac{1}{E_{\nu}}\right),
\end{equation}
where $E_\mathrm{r}$ is the electronic recoil energy and $E_{\nu}$ is the energy of the neutrino. Note that Eq.~(\ref{eq:mu_xsec}) assumes free electrons; small corrections need to be made for the electron binding energies at $O$(keV) energies. 

We search for an anomalous magnetic moment using solar neutrinos, predominantly those from the proton-proton ($pp$) reaction~\cite{Bahcall_2004}. The expected energy spectrum for \mbox{$\mu_\nu=7\times 10^{-11}\,\mu_B$} is shown in Fig.~\ref{fig:signals} (right), which was calculated by folding the expected solar neutrino flux \cite{Bahcall_2004} with Eq.~(\ref{eq:mu_xsec}) and applying a step-function approximation to account for the electron binding energies. In the energy range considered here, this approximation agrees well with more detailed calculations~\cite{numu_theory_xsec}. Note that this signal would be added to the SM neutrino elastic scattering spectrum, which we treat as a background as described in Sec.~\ref{subsec:background}.

\subsection{Bosonic Dark Matter}\label{subsec:bdm}
Axion-like particles (ALPs), like QCD axions, are pseudoscalar bosons, but with decay constant and particle mass (Eq.~(\ref{eq:mass_scale})) decoupled from each other and instead taken as two independent parameters. This decoupling allows for ALPs to take on higher masses than QCD axions; however, it also implies that ALPs do not solve the strong CP problem.

ALPs are viable dark matter candidates~\cite{wisp}, and could be absorbed in XENON1T via the axioelectric effect~(Eq.~(\ref{eq:pseudo_xsec})) like their QCD counterparts. Assuming ALPs are \mbox{non-relativistic} and make up all of the local dark matter (density \mbox{$\rho\sim0.3$\,GeV/cm$^3$}~\cite{de_Salas_2019}), the expected signal is a \mbox{mono-energetic peak} at the rest mass of the particle, $m_\mathrm{a}$, with an event rate given by (see~\cite{Pospelov2008a, Arisaka2013})
\begin{equation}
R \simeq \frac{1.5 \times 10^{19}}{A} g_{\mathrm{ae}}^2 \left(\frac{m_\mathrm{a}}{\text{keV}/c^2}\right) \left(\frac{\sigma_\mathrm{pe}}{\text{b}}\right)\text{kg}^{-1}\text{d}^{-1}, 
\label{eq:pseudo_rate}
\end{equation}
where $A$ is the average atomic mass of the detector medium ($A\approx131\,\mathrm{u}$ for xenon). The rate coefficient from our calculation is consistent with \cite{Bloch:2016sjj} for the dark matter density used in this work.

In addition to the pseudoscalar ALPs, XENON1T is also sensitive to vector bosonic dark matter, of which dark photons are a common example. Dark photons can couple weakly with SM photons through kinetic mixing~\cite{Galison} and be absorbed with cross section $\sigma_\mathrm{V}$ given by~\cite{theory_haipeng_2014}
\begin{equation}
\sigma_\mathrm{V} \simeq \frac{\sigma_\mathrm{pe}}{\beta}\kappa^2,
\label{eq:vector_xsec}
\end{equation}
where $\sigma_\mathrm{pe}$, $\alpha$, and $\beta$ are the same as in Eq.~(\ref{eq:pseudo_xsec}), and $\kappa$ parameterizes the strength of kinetic mixing between the photon and dark photon. Similarly to Eq.~(\ref{eq:pseudo_rate}), by following the calculation in~\cite{Pospelov2008a}, the rate for non-relativistic dark photons in a detector reduces to 
\begin{equation}
R \simeq \frac{4.7 \times 10^{23}}{A} {\kappa}^2 \left(\frac{\text{keV}/c^2}{m_\mathrm{V}}\right) \left(\frac{\sigma_\mathrm{pe}}{\text{b}}\right)\text{kg}^{-1}\text{d}^{-1},
\label{eq:vector_rate}
\end{equation}
where $m_\mathrm{V}$ is the rest mass of the vector boson. Like the pseudoscalar above, absorption of a vector boson would also result in a monoenergetic peak broadened by the energy resolution of the detector, but with a rate that is inversely proportional to the particle mass. The expected spectrum for a 20\,keV/c$^2$ ALP with $g_{\mathrm{ae}}=2\times10^{-13}$ is shown in Fig.~\ref{fig:signals} (right). Vector bosons have the same signature as ALPs, but the rate scales differently with mass (see Eqs. (\ref{eq:pseudo_rate}, \ref{eq:vector_rate})).

\section{Data Analysis}\label{sec:analysis}
This section describes the data-analysis methods employed to search for the aforementioned signals. The event-selection criteria and their overall efficiency, the detection efficiency, as well as the determination of fiducialization and ROI are given in Sec.~\ref{subsec:dataselection}. Sec.~\ref{subsec:background} details each component of our background model, the predictions of which are consistent with the results of a background-only fit to the data. In Sec.~\ref{subsec:stats}, we define the likelihood used for the fitting and discuss the statistical framework.
\subsection{Data Selection} \label{subsec:dataselection}
The data-selection criteria for this search are similar to~\cite{xe1t_sr0_sr1}, with the selections and efficiencies optimized and reevaluated for the different parameter space and extended energy range. For an event to be considered valid, an S1-S2 pair is required. A valid S1 demands coincident signals in at least 3 PMTs, and a 500 photoelectron (PE) threshold is imposed on the S2 size. This S2 threshold is more stringent than that in~\cite{xe1t_sr0_sr1} in order to reject background events originating from radon daughters on the TPC surface~\cite{xe1t_ana_paper_2}. Since signal events are expected to deposit energy only once in the detector, events with multiple interaction sites are removed. A variety of selection criteria are applied to ensure data quality and a correct S1 and S2 pairing, which is detailed in \cite{xe1t_ana_paper_1}. The efficiencies and uncertainties of the selection criteria are estimated in a procedure similar to \cite{xe1t_ana_paper_1}, and the cumulative selection efficiency is determined using an empirical fit of the data. The average cumulative selection efficiency over the (1, 210)\,keV region is ($91.2\pm0.3$)\%.

\begin{figure}[!htbp]
    \centering
    \includegraphics[width=1\linewidth]{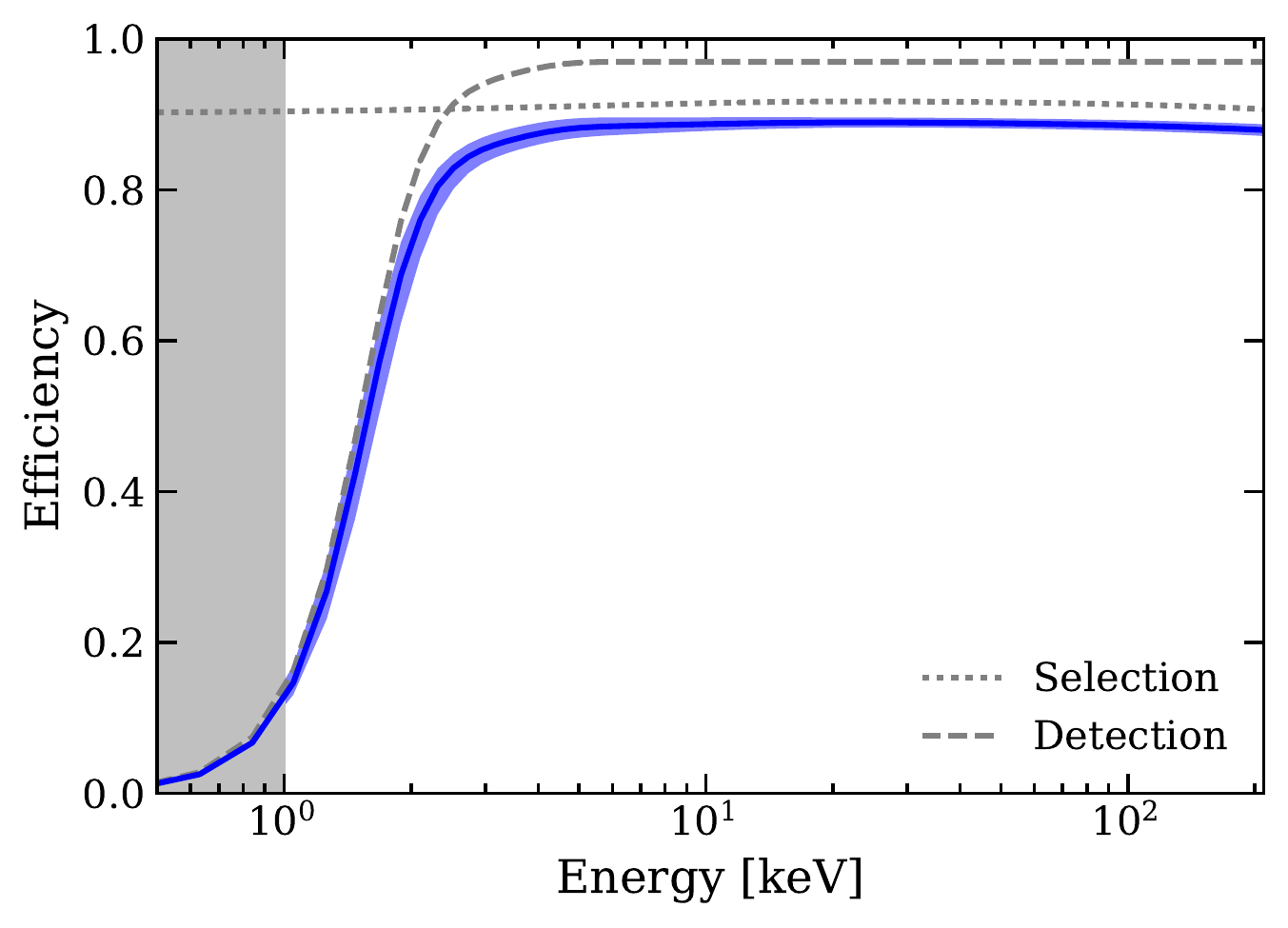}
    \caption{Efficiency as a function of energy. The dashed (dotted) line refers to detection (selection) efficiency, while the blue curve and band illustrate the total efficiency and the associated 1-$\sigma$ uncertainty, respectively. The detection threshold is indicated by the right bound of the gray shaded region.}
    \label{fig:efficiency}
\end{figure}

The combined efficiency of detection and event selection with uncertainties is shown in Fig. \ref{fig:efficiency}. The detection efficiency, dominated by the 3-fold coincidence requirement of S1s, was estimated using both a data-driven method of sampling PMT hits from S1s in the \mbox{20--100\,PE} range and an independent study based on simulation of low-energy S1 waveforms~\cite{xe1t_ana_paper_1}. The  difference between the two methods ($\sim 3\%$ average relative difference in the drop-off region) was considered as a systematic uncertainty. This efficiency was then converted from S1 to reconstructed energy using the detector-response model described in~\cite{xe1t_ana_paper_2}, accounting for additional uncertainties such as the photon yield. The S2 efficiency can be assumed to be unity for the energies considered here~\cite{xe1t_ana_paper_1}. 

Events with energies between (1, 210)\,keV are selected for this search, with the lower bound determined by requiring the total efficiency be larger than 10\%, shown in Fig. \ref{fig:efficiency},  and the upper bound limited by an increasing $\upgamma-$ray background from detector materials, which is difficult to model due to large uncertainties on its spectral shape. While extending the ROI to 210 keV is primarily motivated by the bosonic dark matter search, all profile likelihood fits use this full energy range, as it also allows for better constraints on the background components. The same 1042~kg cylindrical fiducial volume as in \cite{xe1t_firstresult} was used to reduce the surface and material backgrounds. After event selection and strict fiducialization, the surface backgrounds, accidental coincidences, and neutrons make up less than 0.003$\%$ of the total events\,($< 0.3\%$ below 7\,keV), and thus are negligible for this search. Additionally, events within 24 hours from the end of calibration campaigns using injected radioactive sources were removed due to residual source activity. The final effective SR1 live time is 226.9\,days and thus the total exposure is 0.65 tonne-years.

\subsection{Background Model} \label{subsec:background}
\begin{figure}[!htbp]
    \centering
    \includegraphics[width=1\linewidth]{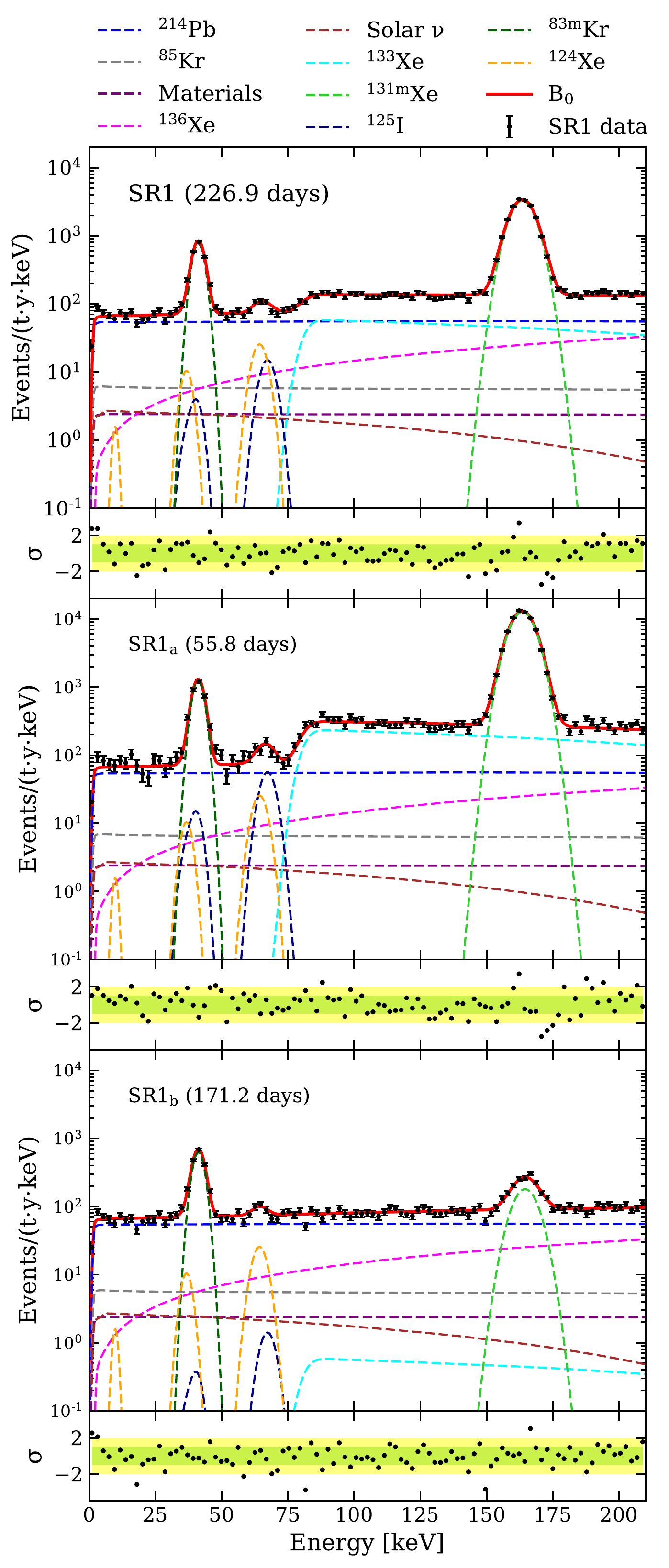}
    \caption{Fit to the SR1 data set using the likelihood framework described in Sec.~\ref{subsec:stats} and the background model $B_0$ in Sec.~\ref{subsec:background}. The top panel shows the entire SR1 spectrum, the sum of the two spectra below it. The middle (bottom) panel shows SR$1_\mathrm{a}$ (SR$1_\mathrm{b}$), which contains more (less) neutron-activated backgrounds. SR$1_\mathrm{a}$ and SR$1_\mathrm{b}$ are fit simultaneously. The light green (yellow) band indicates the 1-$\sigma$ (2-$\sigma$) residuals. The summed fit results are listed in Tab.~\ref{tab:backgrounds}.}
    \label{fig:bg_model}
\end{figure}

Within the (1, 210) keV ROI and the 1042\,kg fiducial volume, ten different components were used to model the background and fit the data, as listed in Tab.~\ref{tab:backgrounds} and illustrated in Fig.~\ref{fig:bg_model}. 

Six components, numbers i--vi in Tab.~\ref{tab:backgrounds}, exhibit continuous energy spectra and were modeled based on either theoretical predictions or GEANT4 Monte Carlo simulations, and the rest are mono-energetic peaks that were modeled as Gaussian functions of known energies and resolution. The spectrum of each background component considers the detector energy resolution and efficiency loss in the same way as the signal model construction in Sec.~\ref{sec:signalmodel}. The rates of the background components are constrained, when possible, by independent measurements and extracted by the fit.

\renewcommand{\arraystretch}{1.2}
\begin{table}[htb]
    \centering
    \begin{tabular}{
            >{\raggedright}m{0.4cm}
    		>{\centering}m{2.4cm} 
    		>{\raggedleft}m{1.3cm}
    	    c
    		>{\raggedright}m{1.3cm} 
    		>{\raggedleft}m{0.8cm} 
    		c 
    		>{\raggedright\arraybackslash}m{0.8cm}
    	}
        \hline\hline
        No.  &  Component  & \multicolumn{3}{c}{Expected Events} & \multicolumn{3}{c}{Fitted Events}   \\

        \hline
i & $^{214}$Pb & \multicolumn{3}{c}{(3450, 8530)} &  7480 & $\pm$ & 160\\
ii & $^{85}$Kr & 890 & $\pm$ & 150  & 773 & $\pm$ & 80\\
iii & Materials & \multicolumn{3}{c}{$323$ (fixed)}  & \multicolumn{3}{c}{$323$ (fixed)} \\
iv & $^{136}$Xe & 2120 &  $\pm$ & 210 & 2150 &  $\pm$ & 120 \\
v & Solar neutrino & 220.7 & $\pm$ & 6.6 & 220.8 & $\pm$ & 4.7\\
vi & $^{133}$Xe & 3900 &  $\pm$ & 410 & 4009 & $\pm$ & 85\\
vii & $^{131\mathrm{m}}$Xe & 23760 & $\pm$ & 640 & 24270 & $\pm$ & 150\\
\multirow{3}{*}{viii} & $^{125}$I (K) & 79 & $\pm$ & 33 & 67 & $\pm$ & 12\\
& $^{125}$I (L) & 15.3 & $\pm$ & 6.5 & 13.1 & $\pm$ & 2.3\\
& $^{125}$I (M) & 3.4 & $\pm$ & 1.5 & 2.94 & $\pm$ & 0.50\\
ix & $^{83\mathrm{m}}$Kr & 2500 & $\pm$ & 250 & 2671 & $\pm$ & 53\\
\multirow{3}{*}{x} & $^{124}$Xe (KK) & 125 & $\pm$ & 50 & 113 & $\pm$ & 24\\
& $^{124}$Xe (KL) & 38 & $\pm$ & 15 & 34.0 & $\pm$ & 7.3\\
& $^{124}$Xe (LL) & 2.8 & $\pm$ & 1.1 & 2.56 & $\pm$ & 0.55\\
        \hline\hline
    \end{tabular}
    \caption{Summary of components in the background model $B_0$ with expected and fitted number of events in the 0.65\,tonne-year exposure of SR1. Both numbers are within the (1, 210) keV ROI and before efficiency correction. See text for details on the various components.}
    \label{tab:backgrounds}
\end{table}

The $\upbeta$ decay of $^{214}$Pb, the dominant continuous background, is present due to $^{222}$Rn emanation into the LXe volume by materials. An additional background comes from intrinsic $^{85}$Kr, which is subdominant due to its removal via cryogenic distillation~\cite{kr_distillation_xe1t, Murra_thesis}. The shape of these spectra, particularly at low energies, can be affected by atomic screening and exchange effects, as well as by nuclear structure~\cite{1992NNDC, Harston92}. The $\upbeta$ decays of $^{214}$Pb and $^{85}$Kr are first forbidden non-unique and first forbidden unique transitions, respectively; however spectra from the IAEA LiveChart (Nuclear Data Services database)~\cite{NDS} 
are based on calculations of allowed and forbidden unique transitions, neither of which includes exchange effects~\cite{2015MO10}. Likewise, models from GEANT4~\cite{Agostinelli:2002hh} include only the screening effect; however, its implementation displays a non-physical discontinuity at low energies~\cite{Hauf,2015MO10}. For this work, we performed dedicated theoretical calculations to account for possible low-energy discrepancies from these effects in $^{214}$Pb and $^{85}$Kr spectra. These calculations are described in detail in Appendix~\ref{sec:appendix}.

The activity of $^{214}$Pb can be constrained using \mbox{\emph{in situ}} measurements of other nuclei in the same decay chain. These constraints, described in~\cite{xe1t_ana_paper_2}, place a lower bound of $5.1\pm0.5$\,$\upmu$Bq/kg (from coincident $^{214}$BiPo) and upper bound of $12.6 \pm 0.8$ $\upmu$Bq/kg ($^{218}$Po $\upalpha$-decays). For this analysis, we leave the normalization of the $^{214}$Pb rate unconstrained and use the fit to extract the activity. The background-only fit results give an event rate of 63.0 $\pm$ 1.3\,events/(tonne$\times$year$\times$keV) (abbreviated as events/(t$\cdot$y$\cdot$keV) for the rest of paper) over the ROI after efficiency correction. With the 11\% branching ratio (from~\cite{WU2009681}) and the spectrum of $^{214}$Pb decay to the ground state (calculated in Appendix~\ref{sec:appendix}), the $^{214}$Pb activity is evaluated to be \mbox{($11.1\pm0.2_\mathrm{stats}\pm1.1_\mathrm{sys} $)\,$\upmu$Bq/kg} throughout SR1 and is well within the upper/lower bounds. The 10\% systematic uncertainty is mainly from the aforementioned branching ratio~\cite{WU2009681}.  

The $^{85}$Kr decay rate is inferred from dedicated measurements of the isotopic abundance of $^{85}$Kr/$^\mathrm{nat}$Kr ($2\times10^{-11}$\,mol/mol) and the $^{\mathrm{nat}}$Kr concentration evolution in LXe~\cite{Lindemann:2013kna}. The same measurements also allow for the time-dependence of the $^{85}$Kr decay rate to be taken into account. The average rate of $^{85}$Kr is $7.4 \pm 1.3$\,events/(t$\cdot$y$\cdot$keV) over the ROI in SR1. 

An additional background arises from \mbox{$\upgamma$ emissions} from radioimpurities in detector materials that induce Compton-scattered electrons; however, this background is subdominant in the ROI due to the strict fiducial volume selection. The rate from materials is constrained by radioassay measurements~\cite{Aprile:2017ilq} and predicted by simulations~\cite{Aprile:2015uzo} to be \mbox{$2.7 \pm 0.3$\,events/(t$\cdot$y$\cdot$keV)}. This background is modeled by a fixed, flat component in the fit.

One of the continuous backgrounds considered was $^{136}$Xe, a 2$\nu\upbeta\upbeta$ emitter intrinsic to xenon. This component has an increasing rate as a function of energy over the ROI. It was constrained in the fit according to the predicted rate and associated uncertainties on (1) a $^{136}$Xe isotopic abundance of \mbox{($8.49\pm0.04_\textrm{stat}\pm0.13_\textrm{sys}$)\%} as measured by a residual gas analyzer \cite{Fieguth_thesis}, (2) the reported half-life~\cite{exo_refined_xe136}, and (3) the calculated theoretical spectrum~\cite{yale2bvv, yalewebsite}.

The first observation of two-neutrino double electron capture (2$\nu$ECEC) of $^{124}$Xe was recently reported using mostly the same SR1 dataset (but different selection cuts) as used in this analysis~\cite{xe1t_dec} and is treated as a background here. In \cite{xe1t_dec} we considered the dominant branching ratio of 2$\nu$ECEC, the capture of two K-shell electrons inducing a peak at 64.3\,keV. It is also possible to capture a K-shell and L-shell electron (36.7\,keV) or two L-shell electrons (9.8\,keV)  with decreasing probability, as calculated in~\cite{dec_branchingratio}. For this analysis, the event selection and consideration of time dependence allow us to include all three peaks in the background model. The predicted rates of the peaks are taken from an updated half-life~\cite{wittweg2020detection} with fixed branching ratios from~\cite{dec_branchingratio}; the overall rate was not constrained in the fit since the half-life was derived from the same dataset.

Three additional backgrounds were included for neutron-activated isotopes: $^{133}$Xe ($\upbeta$), $^{131\mathrm{m}}$Xe (internal conversion (IC)), and $^{125}$I (electron capture (EC)). These isotopes were produced during neutron calibrations and decayed away with half-lives of $O$(10) days. The IC decay of $^{131\mathrm{m}}$Xe produces a mono-energetic peak at 164\,keV~\cite{KHAZOV20062715}, which, along with the other mono-energetic backgrounds, has the same signature as a bosonic dark matter signal. It was well-constrained using its half-life and known dates of neutron calibration. $^{133}$Xe decays to an excited state with a dominant branching ratio and emits an 81\,keV prompt $\upgamma$ upon de-excitation~\cite{KHAZOV2011855}, resulting in a continuous spectrum starting at $\sim$ 75\,keV, given the energy resolution. The rate was also constrained in the fit with prediction obtained using time dependence. The third activated isotope $^{125}$I, a daughter of $^{125}$Xe, decays via EC of K-shell, L-shell, and M-shell with decreasing probability and produces peaks at 67.3\,keV, 40.4\,keV, and 36.5\,keV, respectively~\cite{TabRad_v6}. Similar to $^{124}$Xe 2$\nu$ECEC, all three peaks of $^{125}$I EC are included in the background model with the fixed branching ratios from~\cite{TabRad_v6}. The $^{125}$I contribution was constrained using a model based on the time evolution of $^{125}$Xe throughout SR1, as detailed in~\cite{xe1t_dec}. 

During SR1, a background from $^{83\mathrm{m}}$Kr (IC) was present due to a trace amount of $^{83}$Rb (EC, T$_{1/2} \sim$~86\,days) in the xenon recirculation system, which presumably was caused by a momentary malfunction
of the source valve and confirmed using half-life measurements. $^{83\mathrm{m}}$Kr decays via a two-step scheme (second step T$_{1/2}\sim 154$\,ns)~\cite{manalaysay2010} resulting in many of these events being removed by the \mbox{multi-site} selections mentioned in Sec.~\ref{subsec:dataselection}; however, due to the short half-life of the second step, these decays are often unresolved in time and hence contribute as a mono-energetic peak at 41.5\,keV. This component was also constrained using a time-evolution model.

Elastic scattering of solar neutrinos off electrons is expected to contribute subdominantly over the entire ROI. The expected energy spectrum was obtained using the standard neutrino flux in the Large Mixing Angle Mikheyev-Smirnov-Wolfenstein (\mbox{LMA-MSW}) model and cross section given by the SM~\cite{Bahcall_2004, haxton2013}. Based on rate calculations of neutrino-electron scattering in xenon as given in~\cite{solarnu_calc}, a 3\% uncertainty was assigned and used to constrain the solar neutrino rate in the fit. 

We denote the background model described above as $B_0$. This model was used to fit the SR1 data in \mbox{(1, 210)\,keV} by maximizing the likelihood constructed in Sec.~\ref{subsec:stats}. The fit results are consistent with predictions, as summarized in Tab.~\ref{tab:backgrounds}. The best fit of $B_0$ is shown in Fig.~\ref{fig:bg_model}, where the top panel is the full SR1 data set and the bottom two panels are partitions of SR1, which were fit simultaneously to include the temporal information of several backgrounds (see Sec.~\ref{subsec:stats}). This fit gives a background rate of $76 \pm 2$\,events/(t$\cdot$y$\cdot$keV) within the \mbox{(1, 30)\,keV} region after efficiency correction with the associated uncertainty from the fitting. Fig.~\ref{fig:bg_zoom} shows a zoom in (0, 30)\,keV region of Fig.~\ref{fig:bg_model} with a finer binning.

In Sec.~\ref{sec:results} we raise the possibility of an additional background component, the $\upbeta$ decay of tritium, that we did not include while constructing the background model. A validated $\upbeta-$decay spectrum from the IAEA LiveChart~\cite{NDS,Simpson} was used for the $^3$H model, as described in Appendix~\ref{sec:appendix}. We treat the possible tritium contribution separately from $B_0$ for reasons discussed in Sec.~\ref{subsec:tritium}.

\subsection{Statistical Method} \label{subsec:stats}
An unbinned profile likelihood method is employed in this analysis. The likelihood is constructed as

\begin{align}
\Large{\mathcal{L}}(\mu_s, \boldsymbol{\mu_b}, \boldsymbol{\theta}) = &\mbox{ Poiss}(N|\mu_{tot}) \nonumber \\
&\times \prod^{N}_{i} \left(\sum_j \frac{\mu_{b_j}}{\mu_\mathrm{tot}}f_{b_j}(E_i, \boldsymbol{\theta}) + \frac{\mu_s}{\mu_\mathrm{tot}}f_s(E_i, \boldsymbol{\theta}) \right) \nonumber \\
& \times\prod_m C_{\mu_m}(\mu_{b_m}) \times \prod_n
C_{\theta_n}(\theta_n), \label{eq:likelihood} \\
\mu_\mathrm{tot} &\equiv \sum_{j} \mu_{b_j} + \mu_s, \nonumber
\end{align}

\noindent where $\mu_s$ and $\boldsymbol{\mu_b}$ are the expected total signal and background events. Both $\boldsymbol{\mu_b}$ and $\boldsymbol{\theta}$ are nuisance parameters, where $\boldsymbol{\theta}$ includes shape parameters for the efficiency spectral uncertainty (see Fig. \ref{fig:efficiency}), as well as peak location uncertainties, specifically for $^{124}$Xe (3 peaks), $^{83\mathrm{m}}$Kr, and $^{131\mathrm{m}}$Xe. Having largely subdominant event rates, the 3 peak locations from $^{125}$I EC are fixed at their expected positions to save computation time. Index $i$ runs over all observed events with the total number of \mbox{$N$ (=42251\,events)}, and $E_i$ corresponds to the energy of the $i$th event. $f_b$ and $f_s$ are the background and signal probability distribution functions, and index $j$ runs over all the background components. $C_{\mu}$ and $C_{\theta}$ are constraints on the expected numbers of background events and the shape parameters. Index $m$ runs over backgrounds including $^{85}$Kr, solar neutrino, $^{136}$Xe, $^{83\mathrm{m}}$Kr, $^{125}$I, $^{133}$Xe, and $^{131\mathrm{m}}$Xe, while index $n$ is for all six shape parameters. 

Due to time-dependent backgrounds, the SR1 data set is divided into two partitions: SR1$_\mathrm{a}$ consisting of events within 50\,days following the end of neutron calibrations and SR1$_\mathrm{b}$ containing the rest, with effective live times of 55.8 and 171.2\,days, respectively. Including this time information allows for better constraints on the \mbox{time-independent} backgrounds and improves sensitivity to bosonic dark matter, especially as the \mbox{time-dependent} background from $^{133}$Xe impacts a large fraction of its search region. The full likelihood is then given by
\begin{equation}
\large{\mathcal{L}} = \large{\mathcal{L}}_\mathrm{a} \times \large{\mathcal{L}}_\mathrm{b},
\end{equation}
\noindent where $\mathcal{L}_\mathrm{a}$ and $\mathcal{L}_\mathrm{b}$ are evaluated using Eq.~(\ref{eq:likelihood}) in each partition. Nuisance parameters that do not change with time, along with all of the signal parameters, are shared between the two partitions. The constant nuisance parameters are:
\begin{itemize}
    \item the efficiency parameter, which is dominated by detection efficiency and does not change with time.
    \item The $^{214}$Pb component, which was determined to have a constant rate in time using detailed studies of the $\upalpha$-decays of the $^{222}$Rn and $^{218}$Po as well as the coincidence signature of $^{214}$Bi and $^{214}$Po.
    \item The solar neutrino rate, which would vary by $\sim$3\,\% between the two partitions on account of Earth's orbit around the Sun. This is ignored due to the subdominant contribution from this source.
    \item The decay rates of the intrinsic xenon isotopes $^{136}$Xe and $^{124}$Xe, as well as the Compton continuum from materials.
\end{itemize}
The remaining parameters all display time dependencies that are modeled in the two partitions. 

The test statistic used for the inference is defined as
\begin{equation}
q(\mu_s) = -2 \mbox{ln}\frac{\mathcal{L}(\mu_s, \hat{\hat{\boldsymbol{\mu}}}_b, \hat{\hat{\boldsymbol{\theta}}})}{\mathcal{L}(\hat{\mu}_s, \hat{\boldsymbol{\mu}}_b, \hat{\boldsymbol{\theta}})},
\label{eq:teststats}
\end{equation}

\noindent where $(\hat{\mu}_s, \hat{\boldsymbol{\mu}}_b, \hat{\boldsymbol{\theta}})$ is the overall set of signal and nuisance parameters that maximizes $\mathcal{L}$, while $\mathcal{L}(\mu_s, \hat{\hat{\boldsymbol{\mu}}}_b, \hat{\hat{\boldsymbol{\theta}}})$ is the maximized $\mathcal{L}$ by profiling nuisance parameters with a specified signal parameter $\mu_s$. The statistical significance of a potential signal is determined by $q(0)$. For the neutrino magnetic moment and bosonic dark matter searches, a modified Feldman-Cousins method in~\cite{Mora_2019} was adopted in order to derive 90\% C.L.~bounds with the right coverage. We report an interval instead of an upper limit if the global significance exceeds 3\,$\sigma$. For bosonic dark matter this corresponds to 4\,$\sigma$ local significance on account of the look-elsewhere effect, which is not present for the neutrino magnetic moment search. The 3\,$\sigma$ significance threshold only serves as the transition point between reporting one- and two-sided intervals, and was decided prior to the analysis to ensure correct coverage. A two-sided interval does not necessarily indicate a discovery, which in particle physics generally demands a $5 \sigma$ significance and absence of compelling alternate explanations.

Since the solar axion search is done in the space of $g_{\mathrm{ae}}$, $g_{\mathrm{ae}}g_\mathrm{a\upgamma}$, and $g_{\mathrm{ae}}g_{\mathrm{an}}^\mathrm{eff}$, we extend its statistical analysis to three dimensions. For this search, we use a standard profile likelihood construction where the true 90th-percentile of the test statistic (Eq.~(\ref{eq:teststats})) was evaluated at several points on a three-dimensional grid and interpolated between points to define a 3D `critical' volume of true 90-percent threshold values. By construction, the intersection of this volume with the test statistic $q(g_\mathrm{ae},g_{\mathrm{ae}}g_\mathrm{a\upgamma}, g_{\mathrm{ae}}g_{\mathrm{an}})$ defines a three-dimensional 90\% C.L. volume in the space of the three axion parameters. In Sec.~\ref{sec:results} we report the \mbox{two-dimensional} projections of this volume, found by profiling over the third respective signal component.

\section{Results}
\label{sec:results}
\begin{figure}[!htbp]
    \centering
    \includegraphics[width=1\linewidth]{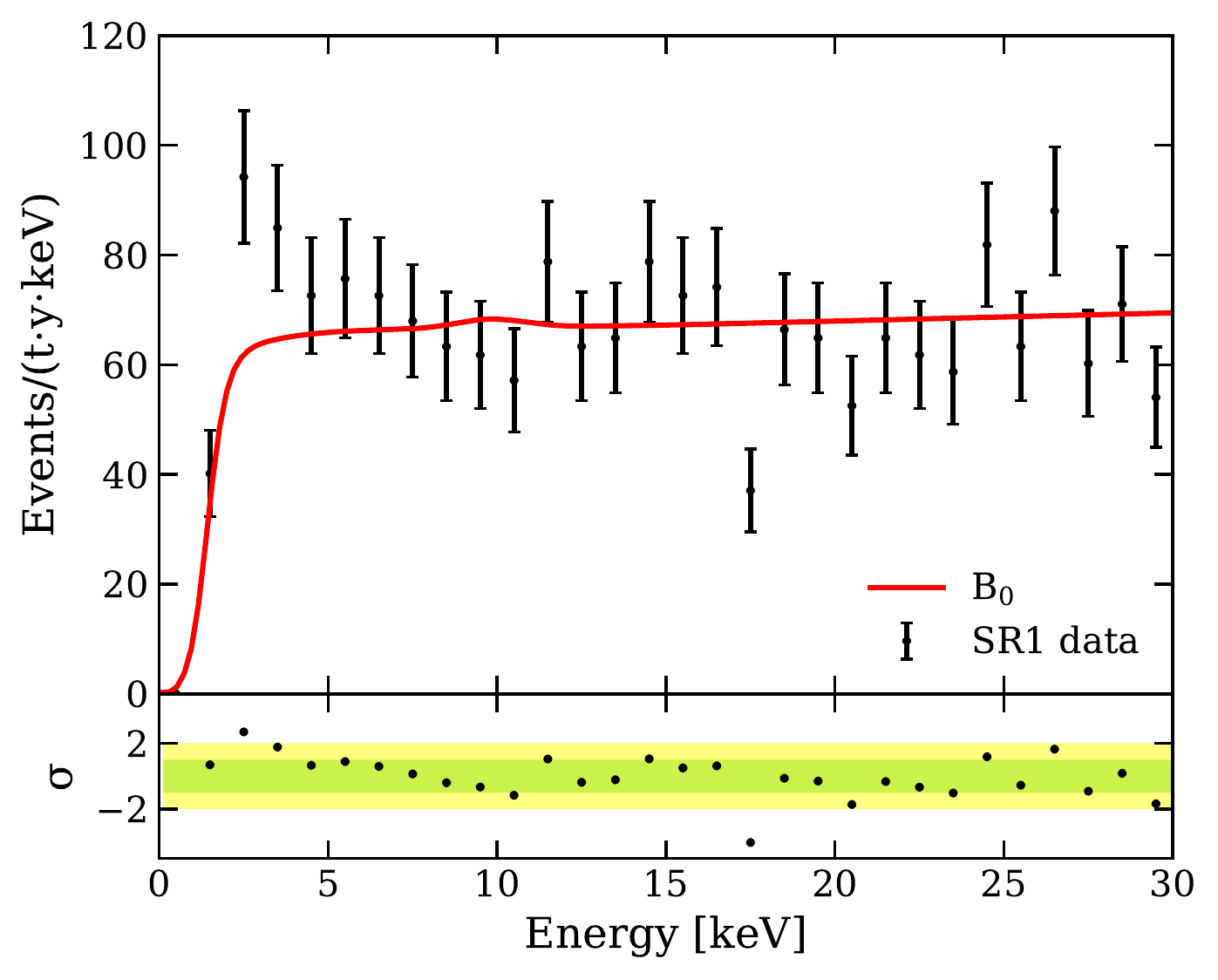}
    \caption{A zoomed-in and re-binned version of Fig.~\ref{fig:bg_model} (top), where the data display an excess over the background model $B_0$. In the following sections, this excess is interpreted under solar axion, neutrino magnetic moment, and tritium hypotheses. } 
    \label{fig:bg_zoom}
\end{figure}

When compared to the background model $B_0$, the data display an excess at low energies, as shown in Fig.~\ref{fig:bg_zoom}. The excess departs slightly from the background model near 7\,keV, rises with decreasing energy with a peak near \mbox{2--3\,keV}, and then subsides to within $\pm$1\,$\sigma$ of the background model near 1--2\,keV. Within this reference region of 1--7\,keV, there are 285 events observed in the data compared to an expected \mbox{$232 \pm 15$ events} from the background-only fit, a $3.3\,\sigma$ Poissonian fluctuation. Events in this energy region are uniformly distributed in the fiducial volume. The temporal distribution of these events are discussed in Sec.~\ref{subsec:sr2}.

\begin{figure}[!htbp]
    \centering
    \includegraphics[width=1\linewidth]{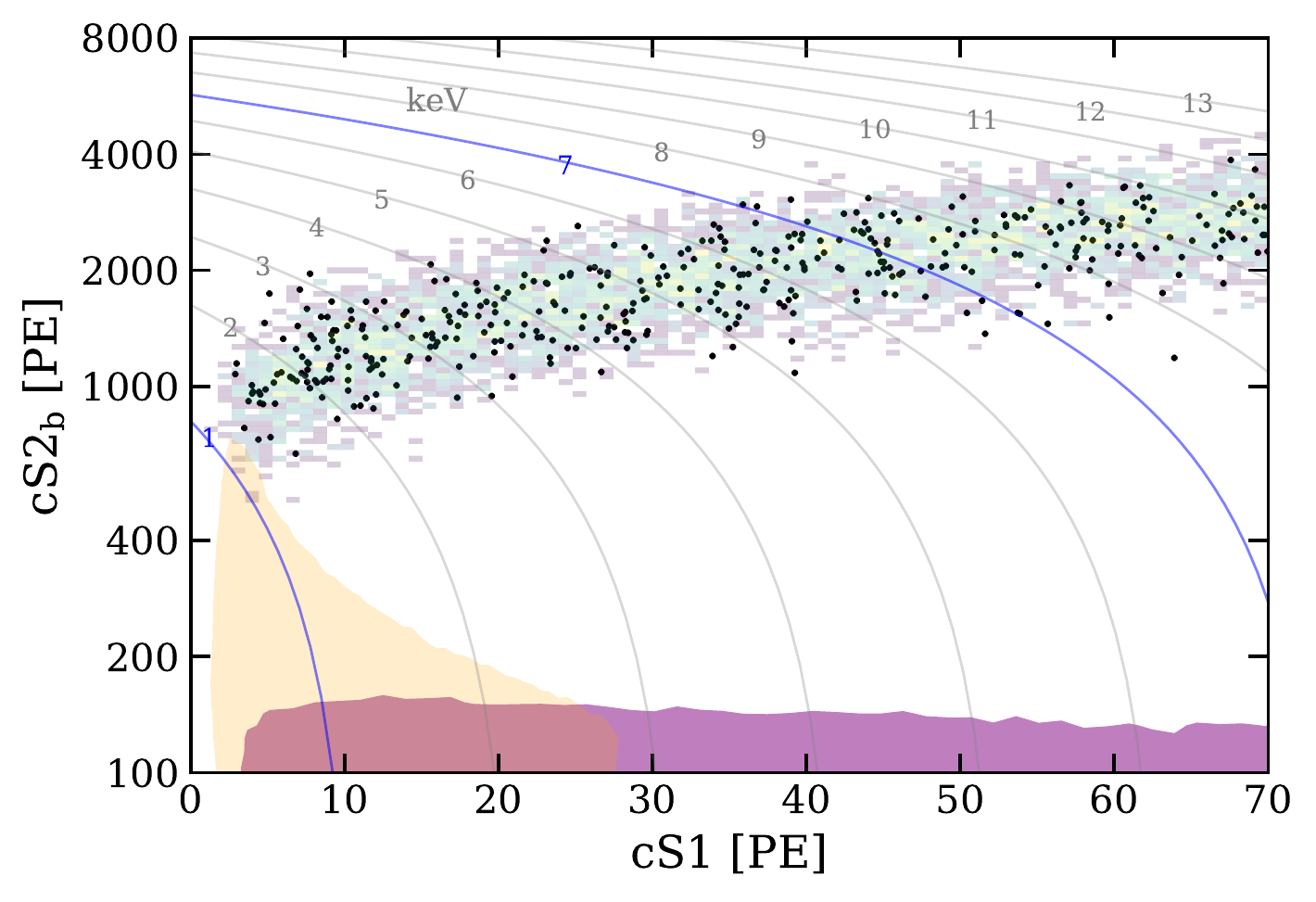}
    \caption{Distribution of low energy events (black dots) in the (cS1, cS$2_b$) parameter space, along with the expected \mbox{surface (purple)} and AC (orange) backgrounds (1\,$\sigma$ band).  $^{220}$Rn \mbox{calibration} events are also shown (density map). All the distributions are within the one-tonne fiducial volume. Gray lines show isoenergy contours for electronic recoils, where 1 and 7\,keV contours, the boundaries of the reference region, are highlighted in blue.} 
    \label{fig:ac_surface}
\end{figure}

Several instrumental backgrounds and systematic effects were excluded as possible sources of the excess. Accidental coincidences (AC), an artificial background from detector effects, are expected to be spatially uniform, but are tightly constrained to have a rate of \mbox{$<1$ event/(t$\cdot$y$\cdot$keV)} based on the rates of lone signals in the detector, i.e., S1s (S2s) that do not have a corresponding S2 (S1)~\cite{xe1t_ana_paper_2}. Surface backgrounds have a strong spatial dependence~\cite{xe1t_ana_paper_2} and are removed by the fiducialization (1.0 tonne here vs. 1.3 tonnes in \cite{xe1t_sr0_sr1}, corresponding to a radial distance from the TPC surface of $\gtrsim$ 11\,cm) along with the stricter S2 threshold cut. Both of these backgrounds also have well-understood signatures in the (cS1, cS2$_\mathrm{b}$) parameter space that are not observed here, as shown in Fig. \ref{fig:ac_surface}.

\begin{figure}[!htbp]
    \centering
    \includegraphics[width=1\linewidth]{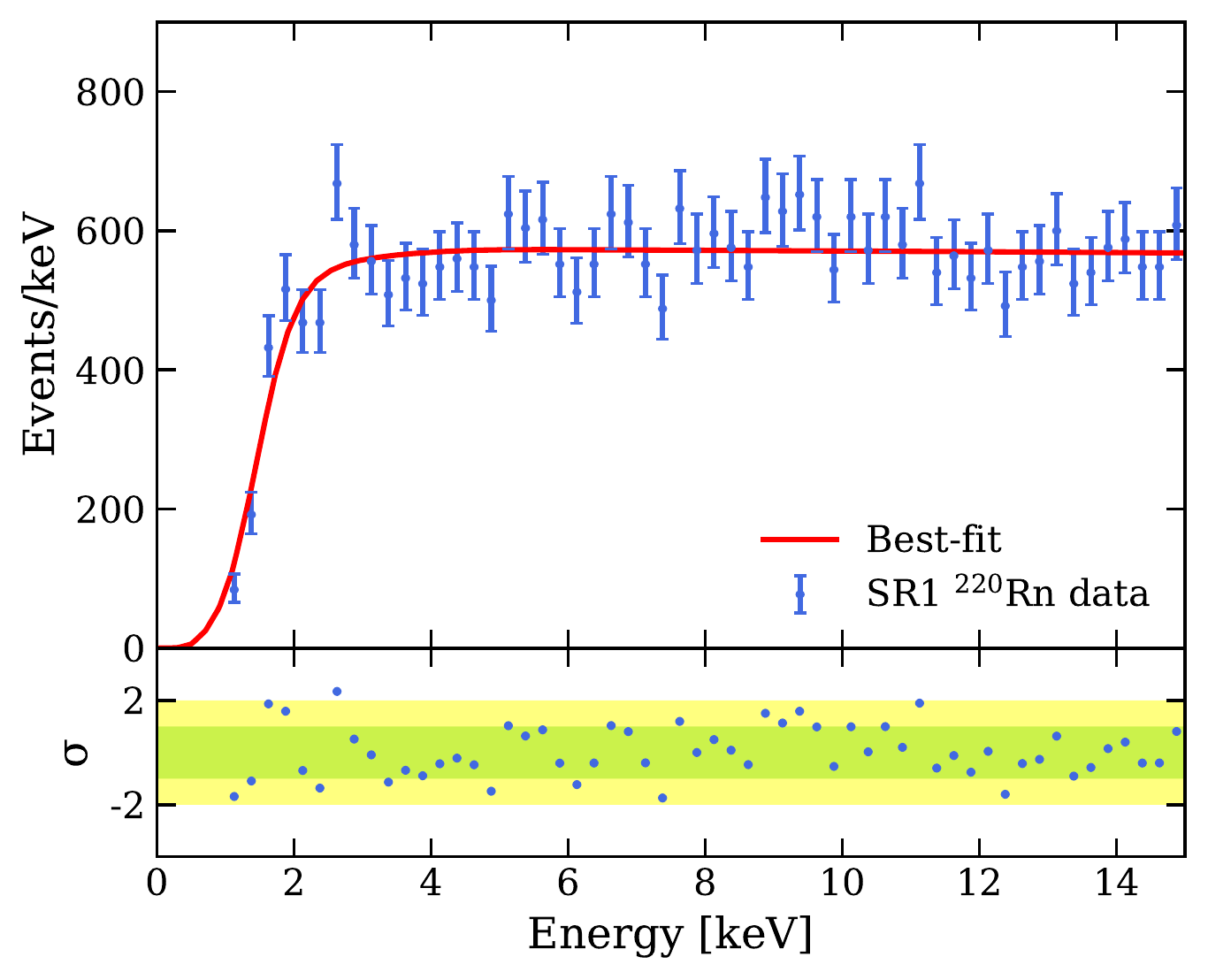}
    \caption{Fit to $^{220}$Rn \textit{calibration} data with a theoretical $\upbeta$-decay model (see Appendix~\ref{sec:appendix}) and the efficiency nuisance parameter, using the same unbinned profile likelihood framework described in Sec.~\ref{subsec:stats}. This fit suggests that the efficiency shown in Fig.~\ref{fig:efficiency} describes well the expected spectrum from $^{214}$Pb, the dominant background at low energies.}   
    \label{fig:rn_220}
\end{figure}

The detection and selection efficiencies were verified using $^{220}$Rn calibration data. The $\upbeta$ decay of $^{212}$Pb, a daughter of $^{220}$Rn, was used to calibrate the ER response of the detector, and thus allows us to validate the efficiency modeling with a high-statistics data set. Similarly to $^{214}$Pb, the model for $^{212}$Pb was calculated to account for atomic screening and exchange effects, as detailed in Appendix~\ref{sec:appendix}. A fit to the $^{220}$Rn data with this model and the efficiency parameter described in Sec.~\ref{subsec:stats} is shown in Fig.~\ref{fig:rn_220} for a 1-tonne fiducial volume, where good agreement is observed (\mbox{p-value $= 0.50$}). Additionally, the S1 and S2 signals of the low-energy events in background data were found to be consistent with this $^{220}$Rn data set, as shown in Fig.~\ref{fig:ac_surface}. This discounts threshold effects and other mismodeling\,(e.g., energy reconstruction) as possible causes for the excess observed in Fig.~\ref{fig:bg_zoom}.

Uncertainties in the theoretical background models were considered, particularly for the dominant $^{214}$Pb component. More details can be found in Appendix~\ref{sec:appendix}, but we briefly summarize them here. A steep rise in the spectrum at low energies could potentially be caused by exchange effects in $\upbeta$-decay emission; however, this component is accurate to within 1\% and therefore negligible with respect to the observed excess. The remaining two components, namely the endpoint energy and nuclear structure, tend to shift the entire $\upbeta$ distribution, rather than cause steep changes over a range of $\sim10$\,keV. Conservatively, the combined uncertainty from these two components is $+6\%$ in the 1--10\,keV region, as described in the Appendix~\ref{sec:appendix}. In comparison, a +50\% uncertainty at 2--3\,keV on the calculated $^{214}$Pb spectrum, as constrained by the higher energy component, would be needed to make up the excess.

We also considered backgrounds that might in principle be present in trace amounts. First, low-energy X-rays from $^{127}$Xe EC, as seen in \cite{lux_solar_axion_2017} and \cite{pandax_solar_axion_2017}, are ruled out for a number of reasons. $^{127}$Xe is produced from cosmogenic activation at sea level; given the short half-life of 36.4\,days and the fact that the xenon gas was underground for $O(\mathrm{years})$ before the operation of XENON1T, $^{127}$Xe would have decayed to a negligible level. Indeed, high-energy $\upgamma$s that accompany these \mbox{X-rays} were not observed, and with their $O(\mathrm{cm})$ mean free path in LXe they could not have left the $O(\mathrm{m})$-sized TPC undetected. For these reasons, we conclude that $^{127}$Xe was no longer present during SR1.

Another potential background is $^{37}$Ar, which decays via EC to the ground state of $^{37}$Cl, yielding a 2.82\,keV peak with a 0.90 branching ratio~\cite{ar37_energy}. It was considered by the LUX collaboration as a background to explain a possible excess rate at \mbox{$\sim$ 3\,keV} in their data~\cite{Akerib:2015rjg}. Its ingress was hypothesized to come from either from an initial  amount in the xenon gas or from an air leak during operations; however, no definitive conclusion was drawn based on measurements of both the leakage rate and the $^{37}$Ar concentration in air at the experimental site~\cite{Akerib:2018zoq}. We consider the two aforementioned possibilities for the introduction of $^{37}$Ar into the xenon target and place quantitative constraints on each source. 

$^{37}$Ar has a half-life of T$_{1/2}$ = 35.0\,days~\cite{ar37_energy} and a typical abundance in $^{\mathrm{nat}}$Ar of $\sim 10^{-20}$ mol/mol~\cite{Richard_ar37}. Given an initial measured $^{\mathrm{nat}}$Ar concentration of $<$5~ppm in the xenon inventory~\cite{hasterok}, $^{37}$Ar decayed to a negligible level, $<$ 1\,events/(t$\cdot$y), by the start of the XENON1T commissioning phase ($>$~400 days). As with krypton, argon is not removed by the getter in the purification system, although it is removed by online $^{85}$Kr distillation (see Sec.~\ref{subsec:background}). This further suppresses its presence prior to SR1. These factors conclusively rule out the presence of $^{37}$Ar from its initial concentration in the xenon inventory.

With respect to an $^{37}$Ar component from a constant air leak, the similarities between krypton and argon noble gases allow us to use $^{\mathrm{nat}}$Kr to constrain the concentration of $^{37}$Ar in the detector. From frequent measurements using rare gas mass spectrometry (RGMS)~\cite{Lindemann:2013kna} and its natural abundance~\cite{kr_abundance}, the observed increasing concentration of $^{\mathrm{nat}}$Kr of \mbox{$< 1$ ppt/year} gives an upper limit on the leak rate of $\leq$ \mbox{0.9 liter/year} during SR1, following online distillation~\cite{dst_inprep}. We make a conservative assumption that the $^{\mathrm{nat}}$Kr increase is due entirely to a leak (neglecting emanation). 

The air inside the experimental hall at LNGS, supplied from outside of the laboratory and fully exchanged within 2.5 hours, has an $^{37}$Ar concentration of $<$ 3.2 mBq/m$^3$, as determined from measurements taken in July 2020 following the methods in~\cite{ar37_gaschrom,Riedmann_2011}. We set a constraint using a robust upper limit of 5 mBq/m$^3$ for the $^{37}$Ar equilibrium concentration to account for possible seasonal variations~\cite{RadioactivityintheAtmosphere,Ar37conc}. The estimate is further refined after considering the differential leak rates of the two noble gases based on their respective viscosities in air, as well as accounting for the relative volatility of argon in liquid/gaseous xenon. Applying these corrections and conservatively assuming that $^{37}$Ar reached an equilibrium activity by the start of SR1, we find that its expected rate is $<$5.2\,events/(t$\cdot$y). To explain the excess in XENON1T, the $^{37}$Ar rate is required to be \mbox{$\sim$ 65\,events/(t$\cdot$y)}, implying that the deduced upper limit is a factor of 13 too low to account for the excess. This conservative constraint on its presence in SR1 therefore excludes $^{37}$Ar from a constant air leak as an explanation for the excess.

The time dependence of a potential $^{37}$Ar background is discussed further in Sec.~\ref{subsec:sr2}; however no clear trend is observed due to low statistics, and any temporal fluctuations are still constrained by the measured krypton concentrations throughout SR1. Given its short half-life, low measured concentration, and strong constraints from the leak hypothesis, we conclude that $^{37}$Ar cannot make up the excess, although it may be present in the detector at a negligible level. 

We also considered an additional background that has never been observed before in LXe TPCs: the $\upbeta$ emission of tritium\footnote{Tritium in the form of tritiated methane has been used for calibration of LXe TPCs~\cite{xe100_tritium, lux_tritium,pandax2017}, including XENON100, but was not used as a calibration source in XENON1T. Following the XENON100 tritium calibration, neither the xenon gas nor the materials that came into contact with the tritiated methane were used in XENON1T.}, which has a \textit{Q}-value of 18.6\,keV and a half-life of 12.3 years~\cite{Lucas_Unterweger}. Tritium may be introduced from predominantly two sources: cosmogenic activation of xenon during above-ground exposure~\cite{ZHANG201662} and emanation of tritiated water (HTO) and hydrogen (HT) from detector materials due to its cosmogenic and anthropogenic abundance. In contrast to $^{127}$Xe and $^{37}$Ar, the tritium hypothesis cannot be ruled out. In Sec.~\ref{subsec:tritium} we consider several possible mechanisms for the introduction of tritium into the detector and the uncertainties involved in its production and reduction processes in an attempt to estimate its concentration.

\subsection{Tritium Hypothesis}
\label{subsec:tritium}
\begin{figure*}[!htbp]
    \centering
    \includegraphics[width=0.95\textwidth]{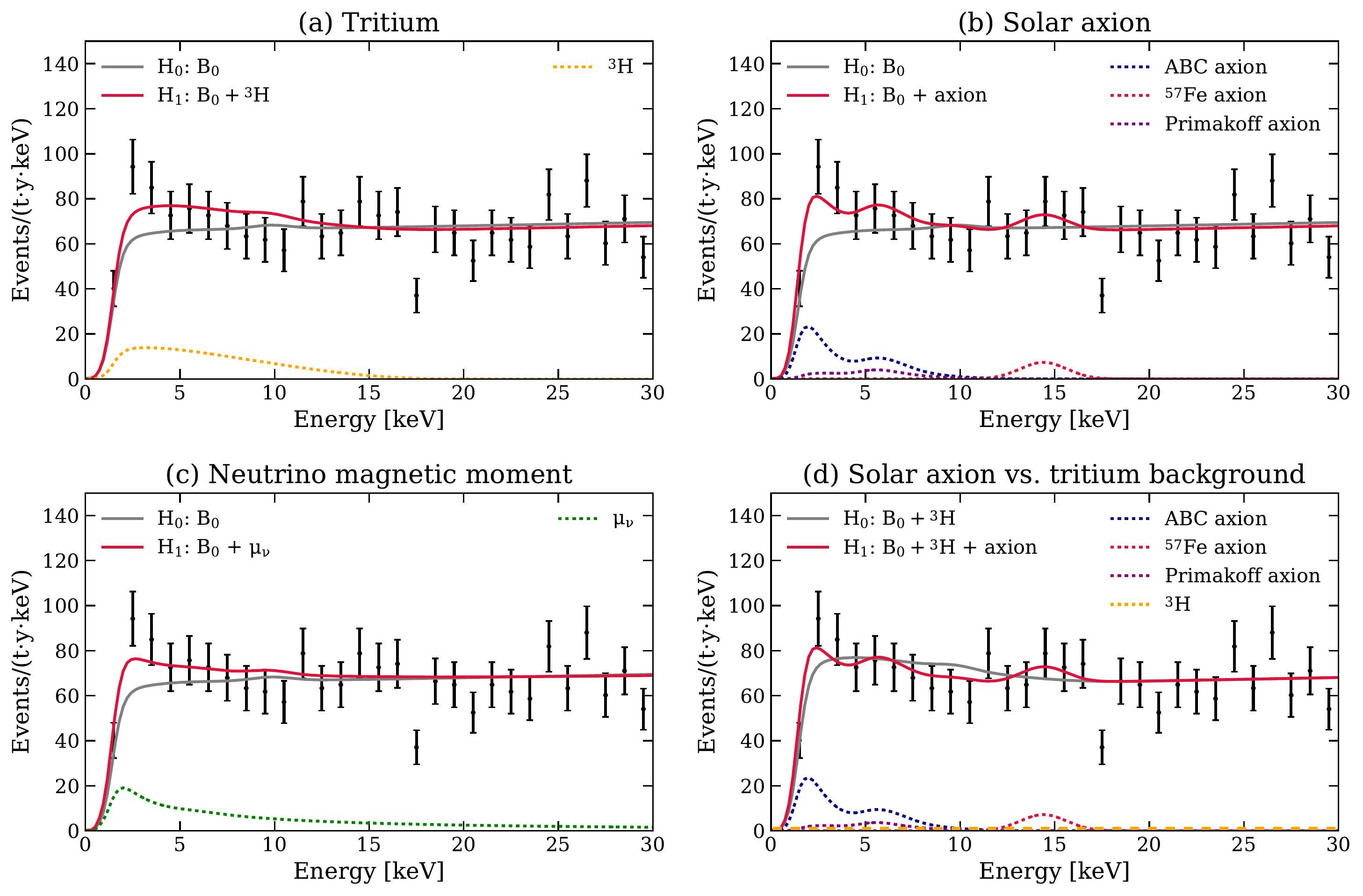}
   \caption{Fits to the data under various hypotheses. The null and alternative hypotheses in each scenario are denoted by gray (solid) and red (solid) lines, respectively. For the tritium (a), solar axion (b), and neutrino magnetic moment (c) searches, the null hypothesis is the background model $B_0$ and the alternative hypothesis is $B_0$ plus the respective signal. Contributions from selected components in each alternative hypothesis are illustrated by dashed lines. Panel (d) shows the best fits for an additional statistical test on the solar axion hypothesis, where an unconstrained tritium component is included in both null and alternative hypotheses. This tritium component contributes significantly to the null hypothesis, but its best-fit rate is negligible in the alternative hypothesis, which is illustrated by the orange dashed line in the same panel.}
\label{fig:low_energy_fits}
\end{figure*}

In order to determine the hypothetical concentration of tritium required to account for the excess, we search for a ${^3\mathrm{H}}$ `signal' on top of the background model $B_0$. When compared to $B_0$, the tritium hypothesis is favored at 3.2\,$\sigma$ and the fitted rate is $159 \pm 51$ events/(t$\cdot$y) (68$\%$\,C.L.), which would correspond to a $^3$H/Xe concentration of $( 6.2 \pm 2.0) \times10^{-25}$\,mol/mol. As tritium is expected to be removed by the xenon purification system, this concentration would correspond to an equilibrium value between emanation and removal. The spectral fits under this hypothesis are illustrated in Fig.~\ref{fig:low_energy_fits}\,(a).

Due to its minute possible concentration, long half-life with respect to our exposure, and the fact that it decays through a single channel, we are unable to confirm the presence of tritium from SR1 data directly. We therefore try to infer its concentration from both initial conditions and detector performance parameters.

A tritium background component from cosmogenic activation of target materials has been observed in several dark matter experiments at rates compatible with predictions~\cite{Amare:2017roa}, although it has never before been detected in xenon. From exposure to cosmic rays during above-ground storage of xenon, we estimate a conservative upper limit on the initial $^3$H/Xe concentration of $< 4 \times 10^{-20}$\,mol/mol, based on GEANT4 activation rates~\cite{ZHANG201662} and assuming saturation activity. At this stage, tritium will predominantly take the form of HTO, given the measured ppm water impurities in the xenon gas and equilibrium conditions~\cite{hasterok, Ishida}. Through xenon gas handling prior to filling the detector (i.e., condensation of H$_2$O/HTO on the walls of the cooled xenon-storage vessel) and purification via a high-efficiency getter with a hydrogen removal unit~\cite{xe1t_instrument,Dobi:2010ai}, we expect the concentration to be reduced to $< 10^{-27}$\,mol/mol, thus reaching negligible levels with respect to the observed excess.

Tritium may also be introduced as HTO and HT via their respective atmospheric abundances. Water and hydrogen, and therefore tritium, may be stored inside materials, such as the TPC reflectors and the stainless steel of the cryostat. This type of source is expected to emanate from detector and subsystem materials at a rate in equilibrium with its removal via getter purification. Tritium can be found in water at a concentration of $(5 - 10) \times 10^{-18}$ atoms of $^3$H for each atom of hydrogen in H$_2$O~\cite{tritium_counting,3H_lngs_2009,HTO_abundance}. Here we assume the same abundance of $^3$H in atmospheric H$_2$ as for water\footnote{Although geographical and temporal HT abundances in the atmosphere vary due to anthropogenic activities, HT that reaches the Earth's surface undergoes exchange to HTO within 5 hours~\cite{HT_history, Mishima_2002}.}. Using the best-fit rate of tritium and the HTO atmospheric abundance, a combined (H$_2$O + H$_2$) impurity concentration of $\gtrsim$ 30\,ppb in the LXe target would be required to make up the excess. Since water impurities affect optical transparency, the high light yield in SR1 indicates an $O(\mathrm{1})$-ppb H$_2$O concentration~\cite{Aprile:2015uzo, elena_LXe_review}, thus implying a maximum contribution from HTO to the $^3$H/Xe concentration of $\sim 1 \times10^{-26}$\,mol/mol. With respect to H$_2$, we currently have no direct or indirect measurements of its concentration in the detector. Instead, we consider that O$_2$-equivalent, electronegative impurities must reach sub-ppb levels in SR1, given the achieved electron lifetime of \mbox{$\sim 650\, \upmu$s} (at 81 V/cm)~\cite{xe1t_sr0_sr1,Aprile:1990pi}. Thus for tritium to make up the excess requires a factor $\sim$ 100 higher H$_2$ concentration than that of electronegative impurities. Under the above assumptions, tritium from atmospheric abundance appears to be an unlikely explanation for the excess. However, we do not currently have measurements of the equilibrium H$_2$ emanation rate in XENON1T, and thus the HT concentration cannot be sufficiently quantified.

In conclusion, possible tritium contributions from cosmogenic activation or from HTO in SR1 appear too small to account for the excess, while it is not possible to infer the concentration of HT. In addition, various factors contribute further to the uncertainty in estimating a tritium concentration within a LXe environment, such as its unknown solubility and diffusion properties, as well as the possibility that it may form molecules other than HT and HTO. Since the information and measurements necessary to quantify the tritium concentration are not available, we can neither confirm nor exclude it as a background component. Therefore, we report results using the background model $B_0$, and then summarize how our results would change if tritium were included as an unconstrained background component. All reported constraints are placed with the validated background model~$B_0$~(i.e., without tritium).

\subsection{Solar Axion Results}\label{subsec:saresult}
We search for ABC, $^{57}$Fe, and Primakoff axions simultaneously. Under this signal model, $B_0$ is rejected at $3.4\,\sigma$, a value determined using toy Monte Carlo methods to account for the three parameters of interest in the alternative hypothesis. A comparison of the best fits under the alternative hypothesis ($B_0 + \mathrm{axion}$) and null hypothesis ($B_0$) can be found in Fig.~\ref{fig:low_energy_fits}\,(b).

\begin{figure}[!htbp]
    \centering
    \raggedright
    \includegraphics[width=1\linewidth]{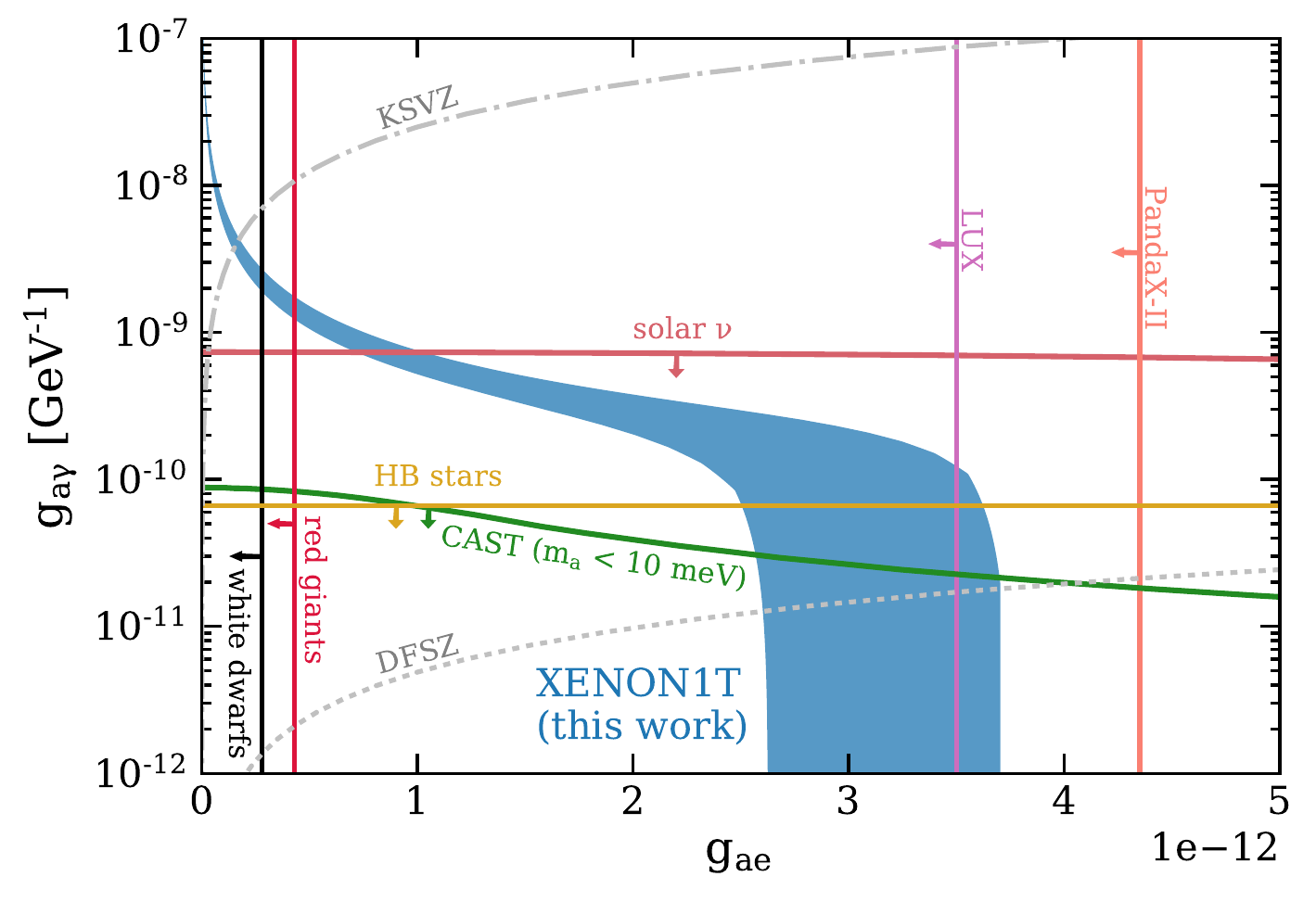}
    \includegraphics[width=1\linewidth]{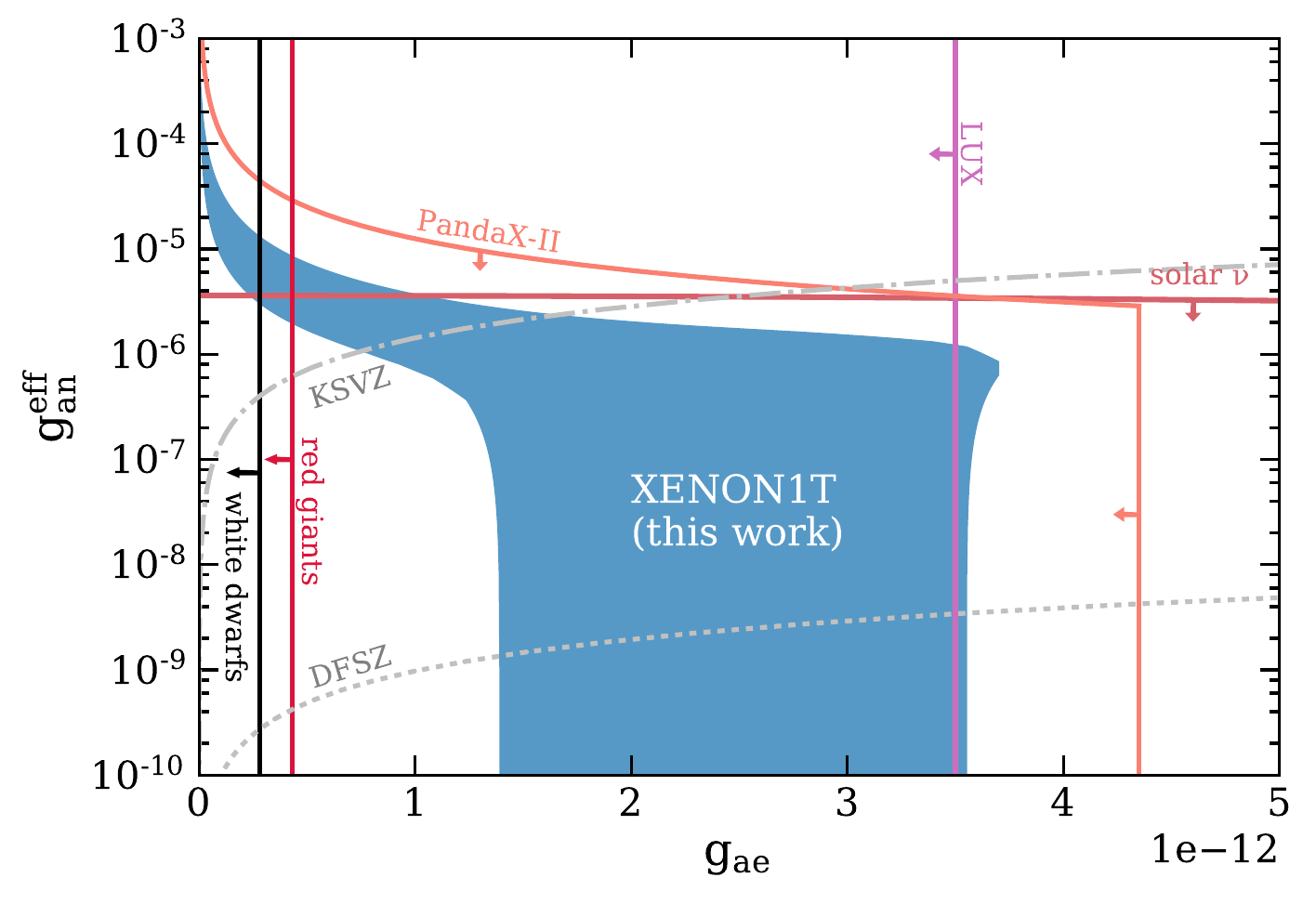}
    \includegraphics[width=1\linewidth]{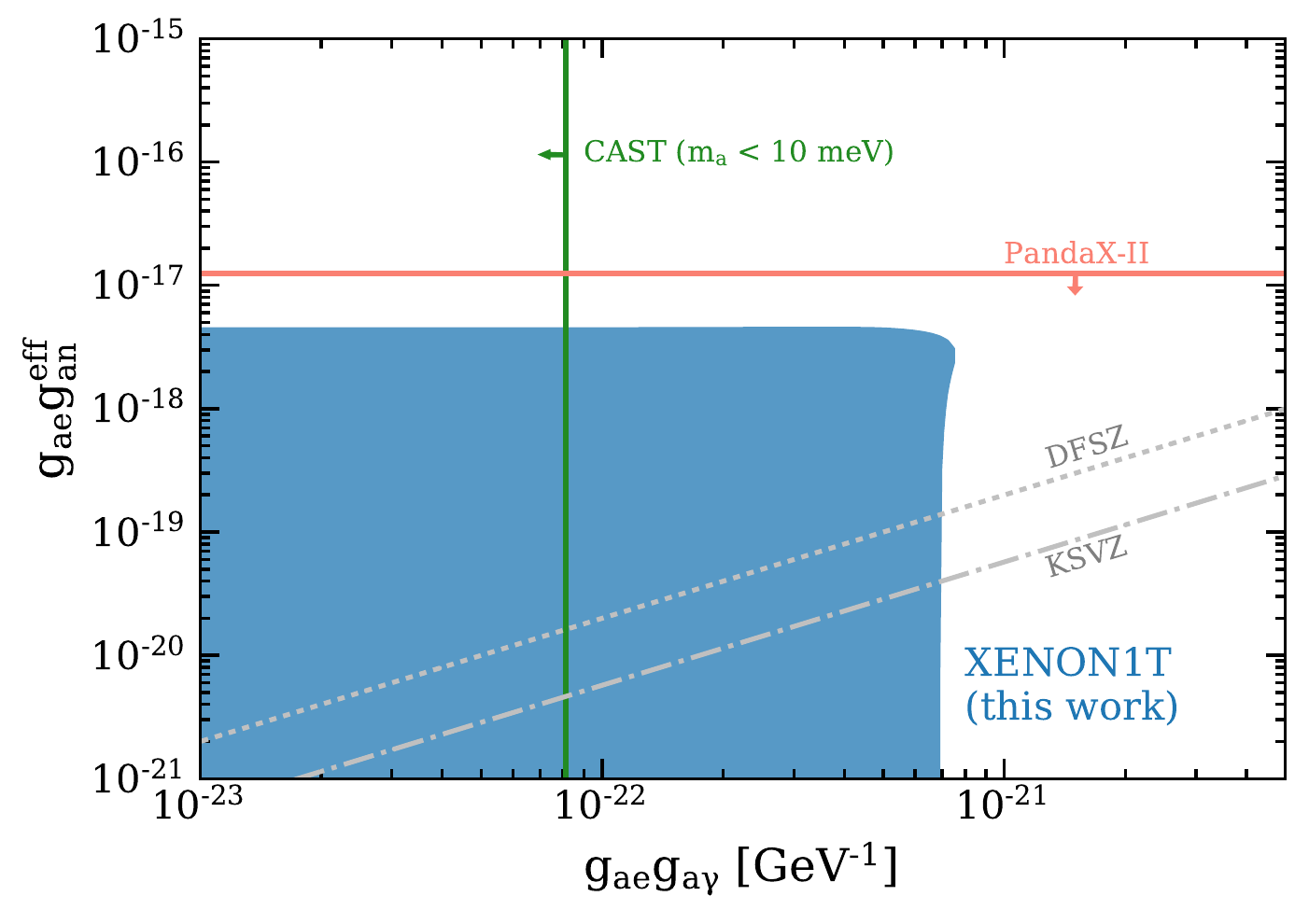}
    \caption{Constraints on the axion-electron $g_{\mathrm{ae}}$, axion-photon $g_\mathrm{a\upgamma}$, and effective axion-nucleon $g_\mathrm{an}^\mathrm{eff}$ couplings from a search for solar axions. The shaded blue regions show the two-dimensional projections of the three-dimensional confidence surface (90\% C.L.) of this work, and hold for $m_\mathrm{a} <$ 100\,eV/c$^2$. See text for more details on the three individual projections. All three plots include constraints (90\% C.L.) from other axion searches, with arrows denoting allowed regions, and the predicted values from the benchmark QCD axion models DFSZ and KSVZ.}
    \label{fig:solax_limit}
\end{figure}

A three-dimensional confidence volume (90\% C.L.) was calculated in the space of $g_{\mathrm{ae}}$ vs. $g_{\mathrm{ae}}g_\mathrm{a\upgamma}$ vs. $g_{\mathrm{ae}}g_\mathrm{an}^\mathrm{eff}$. This volume is inscribed in the cuboid given by
\begin{align*}
    g_{\mathrm{ae}} &< 3.8\times 10^{-12} \\ 
    g_{\mathrm{ae}}g_\mathrm{an}^\mathrm{eff} &< 4.8\times10^{-18} \\
    g_{\mathrm{ae}}g_\mathrm{a\upgamma} &< 7.7\times10^{-22} ~\mathrm{GeV}^{-1}.
\label{eq:boundingbox}
\end{align*}
While easy to visualize, this cuboid is more conservative (it displays over-coverage) than the three-dimensional confidence volume it encloses and does not describe the correlations between the parameters. The correlation information can be found in Fig.~\ref{fig:solax_limit}, which shows the two-dimensional projections of the surface. For the \mbox{ABC--Primakoff} and \mbox{ABC--$^{57}$Fe} projections (Fig.~\ref{fig:solax_limit} top and middle, respectively), $g_{\mathrm{ae}}$ can be easily factored out of the y-axis to plot $g_\mathrm{a\upgamma}$ vs $g_{\mathrm{ae}}$ (top) and $g_\mathrm{an}^\mathrm{eff}$ vs $g_{\mathrm{ae}}$ (middle). This is not as straightforward for the $^{57}$Fe-Primakoff projection (Fig.~\ref{fig:solax_limit} bottom). Also shown in Fig.~\ref{fig:solax_limit} are constraints from other axion searches~\cite{CAST_gae, solar_v_solar_axion_2009, pandax_solar_axion_2017, lux_solar_axion_2017, wdlf_limit, revisiting_axionphoton_hb, redgiant_gae} as well as predicted values from the benchmark QCD models DFSZ and KSVZ.

Fig.~\ref{fig:solax_limit} (top) is extracted from the projection onto the ABC--Primakoff plane. Since the ABC and Primakoff components are both low-energy signals, the 90\% confidence region is anti-correlated in this space and\,---\,due to the presence of the low-energy excess\,---\,suggests either a non-zero ABC component or non-zero Primakoff component. Since our result gives no absolute lower bound on $g_{\mathrm{ae}}$, the limit on the product $g_{\mathrm{ae}}g_\mathrm{a\upgamma}$ cannot be converted into a limit on $g_\mathrm{a\upgamma}$ on its own; i.e., with $g_{\mathrm{ae}}g_\mathrm{a\upgamma}$=$7.6\times10^{-22}\,\mathrm{GeV}^{-1}$, $g_\mathrm{a\upgamma} \rightarrow \infty$ as $g_{\mathrm{ae}} \rightarrow 0$, as shown in Fig.~\ref{fig:solax_limit} (top).  

Fig.~\ref{fig:solax_limit} (middle) is taken from the projection onto the ABC--$^{57}$Fe plane. Unlike the ABC-Primakoff case, these two signals are not degenerate; however, they still display anti-correlated behavior. The reason for this is that the test statistic $q$ (Eq.~(\ref{eq:teststats})) is relatively large with small $g_{\mathrm{ae}}$, meaning small changes in the $^{57}$Fe rate about the best-fit make $q$ cross the 90\% threshold value and thus be excluded by our 90\% confidence volume. There is no statistical significance ($< 1\,\sigma$) for the presence of a 14.4\,keV peak from $^{57}$Fe axions. 

Lastly, Fig.~\ref{fig:solax_limit} (bottom) shows the projection onto the Primakoff-$^{57}$Fe plane, where no correlation is observed. The Primakoff and $^{57}$Fe components are both allowed to be absent as long as there is a non-zero ABC component. This means that, of the three axion signals considered, the ABC component is the most consistent with the observed excess. 

The three projections of Fig.~\ref{fig:solax_limit} can be used to reconstruct the three-dimensional 90\% confidence volume for $g_{\mathrm{ae}}$, $g_{\mathrm{ae}}g_\mathrm{a\upgamma}$, and $g_{\mathrm{ae}}g_\mathrm{an}^\mathrm{eff}$. Due to the presence of an excess at low energy, this volume would suggest either a non-zero ABC component or a non-zero Primakoff component. However, the coupling values needed to explain this excess are in strong tension with stellar cooling constraints~\cite{axion_hunters,revisiting_axionphoton_hb,redgiant_gae,wdlf_limit, DiLuzio:2020wdo}, with the exception of a minute region in the 3D coupling space which corresponds to small $g_{\mathrm{ae}}$ and large $g_\mathrm{an}^\mathrm{eff}$, $g_\mathrm{a\upgamma}$. The CAST constraints~\cite{CAST_gae} as shown are valid for axion masses below 10\,meV/c$^2$ while those from XENON1T and similar experiments hold for all axion masses up to $\sim$ 100\,eV/c$^2$. For an axion mass below 10\,meV/c$^2$, the CAST result prefers the region with large $g_{\mathrm{ae}}$ and small $g_\mathrm{a\upgamma}$; however, there is no tension between the CAST result and this result for higher axion masses ($m_a$ $>$ 250 meV/c$^2$) due to the limited sensitivity of CAST for high-mass axions.

As described above, we cannot exclude tritium as an explanation for this excess. Thus, we report on an additional statistical test, where an unconstrained tritium component was added to the background model $B_0$ and profiled over alongside the other nuisance parameters. In this case, the null hypothesis is the background model plus tritium \mbox{($B_0 + {^3\mathrm{H}}$)} and the alternative includes the three axion signal components \mbox{($B_0 + {^3\mathrm{H}} + \mathrm{axion}$)}, where tritium is unconstrained in both cases. The solar axion signal is still preferred in this test, but its significance is reduced to $2.0\sigma$. The fits for this analysis are shown in Fig.~\ref{fig:low_energy_fits}\,(d). The tritium component is negligible in the alternate best-fit, but its presence allows for a better fit under---and thus a reduced significance of rejecting---the null hypothesis.

\subsection{Neutrino Magnetic Moment Results}\label{subsec:nmmresult}
When compared to the neutrino magnetic moment signal model, the background model $B_0$ is rejected at $3.2\,\sigma$. The best-fits of the null ($B_0$) and alternative (\mbox{$B_0 + \mu_{\nu}$}) hypotheses for this search are shown in Fig.~\ref{fig:low_energy_fits}~(c).

\begin{figure}[!htbp]
    \centering
    \raggedright
    \includegraphics[width=1\linewidth]{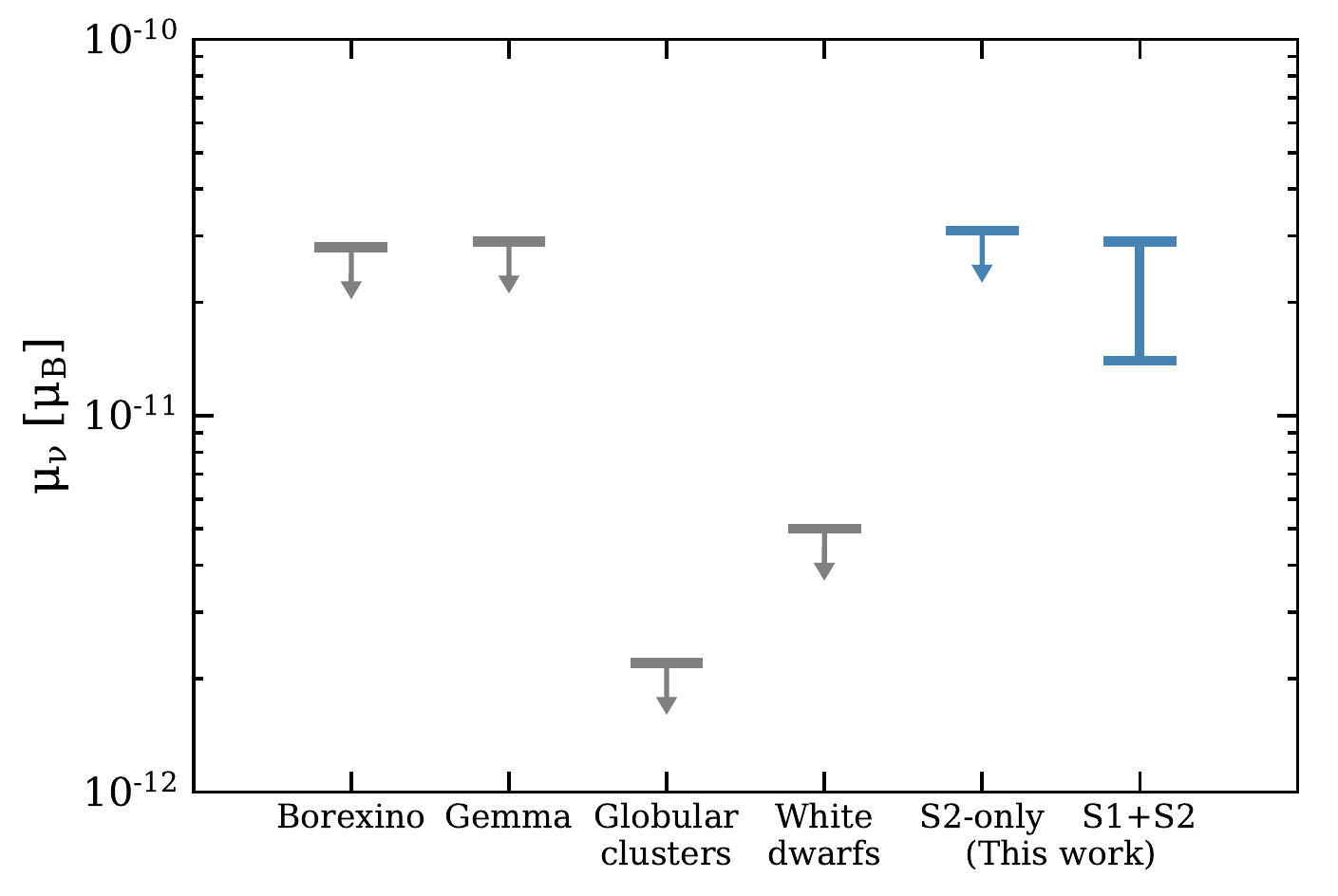}
    \caption{Constraints (90\% C.L.) on the neutrino magnetic moment from this work compared to experiments Borexino~\cite{borexino_nmm} and Gemma~\cite{Gemma_mu}, along with astrophysical limits from the cooling of globular clusters~\cite{redgiants_gae} and white dwarfs~\cite{Corsico_2014}. The constraint from XENON1T using ionization signal only (S2-only) is also shown (see Sec.~\ref{subsec:sr2}). Arrows denote allowed regions. The upper boundary of the interval from this work is about the same as that from Borexino and Gemma. If we interpret the low-energy excess as a neutrino magnetic moment signal, its 90\% confidence interval is in strong tension with the astrophysical constraints.}
    \label{fig:mu_limit}
\end{figure}

The 90\% confidence interval for $\mu_{\nu}$ from this analysis is given by $$\mu_{\nu} \in (1.4,~2.9) \times 10^{-11}\,\mu_B,$$
and is shown in Fig.~\ref{fig:mu_limit} along with the constraints from other searches. The upper boundary of this interval is very close to the limit reported by Borexino \cite{borexino_nmm}, which is currently the most stringent direct detection constraint on the neutrino magnetic moment. Similar to the solar axion analysis, if we infer the excess as a neutrino magnetic moment signal, our result is in strong tension with indirect constraints from analyses of white dwarfs~\cite{Corsico_2014} and globular clusters~\cite{redgiants_gae}. The result is also compatible with the constraint from XENON1T using the S2-only method, which is able to probe a lower energy region and is further discussed in Sec.~\ref{subsec:sr2}. It is important to note that the neutrino flavor does impact the interaction involving the magnetic moment, which in reality is a $3\times3$ matrix due to neutrino mixing. Our result, based on a flavor-insensitive detection of solar neutrinos, is thus directly comparable to Borexino's, but not necessarily to Gemma's (reactor electron anti-neutrinos) or the astrophysical limits (electron neutrinos). 

As in Sec.~\ref{subsec:saresult}, we report on an additional statistical test where an unconstrained tritium component was included in both null and alternative hypotheses. In this test the significance of the neutrino magnetic moment signal is reduced to 0.9\,$\sigma$ with the presence of a tritium background.

This is the most sensitive search to date for an enhanced neutrino magnetic moment with a dark matter detector, and suggests that this beyond-the-SM signal be included in the physics reach of other dark matter experiments.

\subsection{Bosonic Dark Matter Results}
For bosonic dark matter, we iterate over (fixed) masses between 1 and 210\,keV/c$^2$ to search for peak-like excesses. The trial factors to convert between local and global significance were extracted using toy Monte Carlo methods. While the excess does lead to looser constraints than expected at low energies, we find no global significance over $3\,\sigma$ for this search under the background model $B_0$.  We thus set an upper limit on the couplings $g_{\mathrm{ae}}$ and $\kappa$ as a function of particle mass. 

\begin{figure}[!htbp]
    \centering
    \includegraphics[width=1\linewidth]{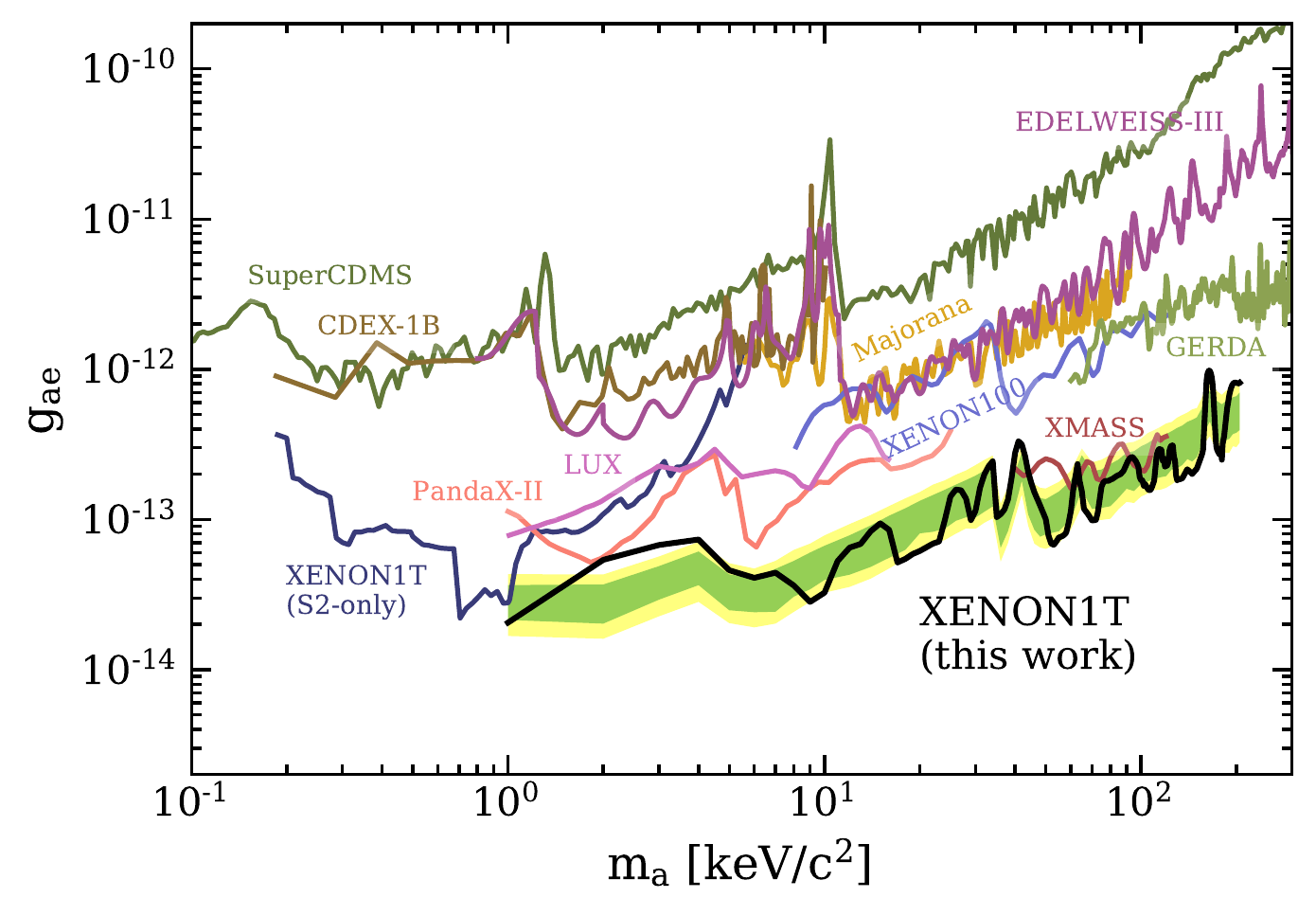}
    \raggedright
    \includegraphics[width=1\linewidth]{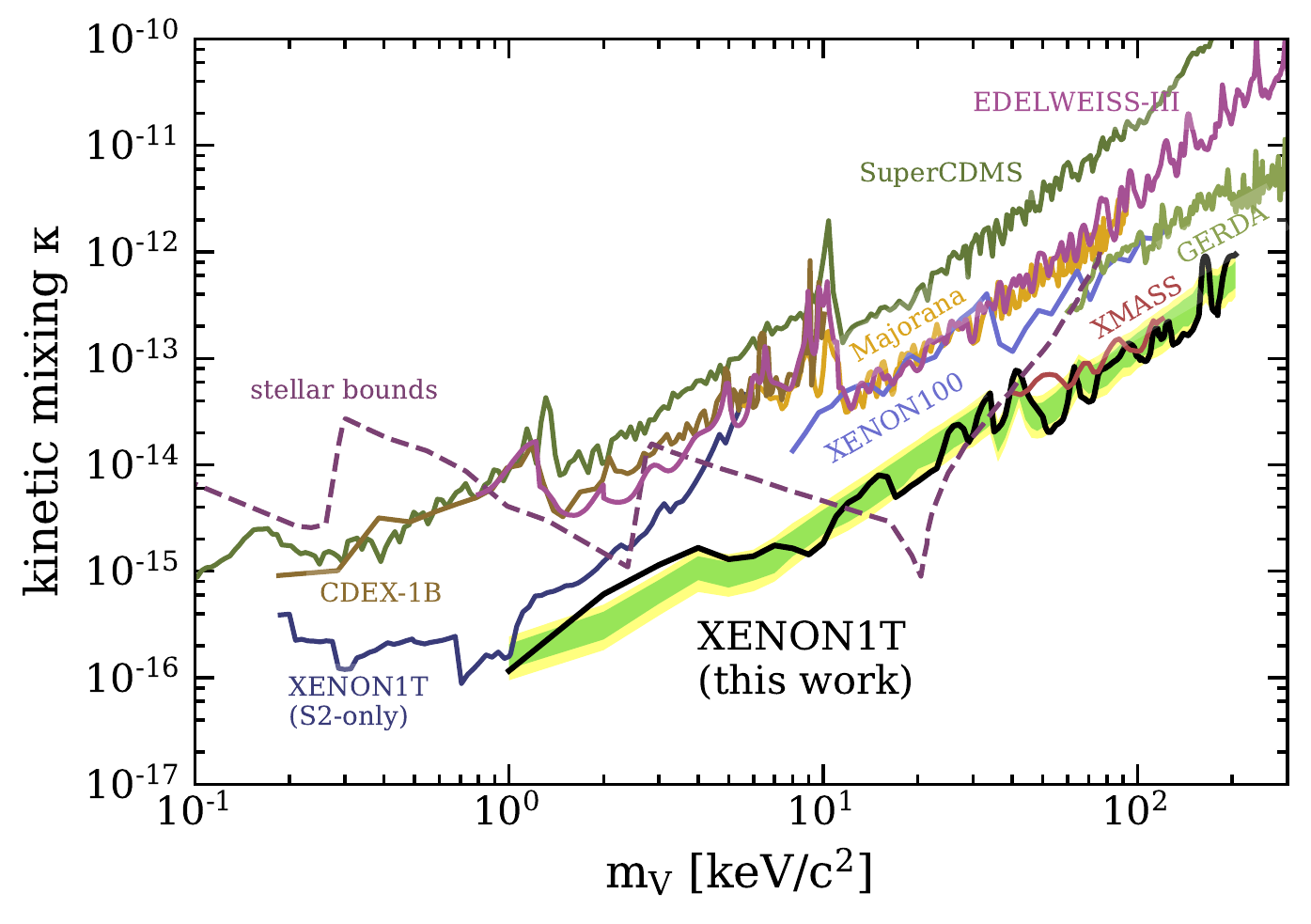}
    \caption{Constraints on couplings for bosonic pseudoscalar ALP (top) and vector (bottom) dark matter, as a function of particle mass. The XENON1T limits (90\% C.L.) are shown in black with the expected 1 (2)\,$\sigma$ sensitivities in green (yellow). Limits from other detectors or astrophysical constraints are also shown for both the pseudoscalar and vector cases~\cite{cdex2019, supercdms2019, 1t_s2only, xmass_2018, EDELWEISS_2018, xe100_2017, xe100_2017, lux_solar_axion_2017, pandax_solar_axion_2017, theory_haipeng_2014,GERDA_SW,majorana}.}
    \label{fig:dp_alps_limit}
\end{figure}

\begin{figure*}[!htbp]
    \centering
    \raggedright
    \includegraphics[width=0.95\textwidth]{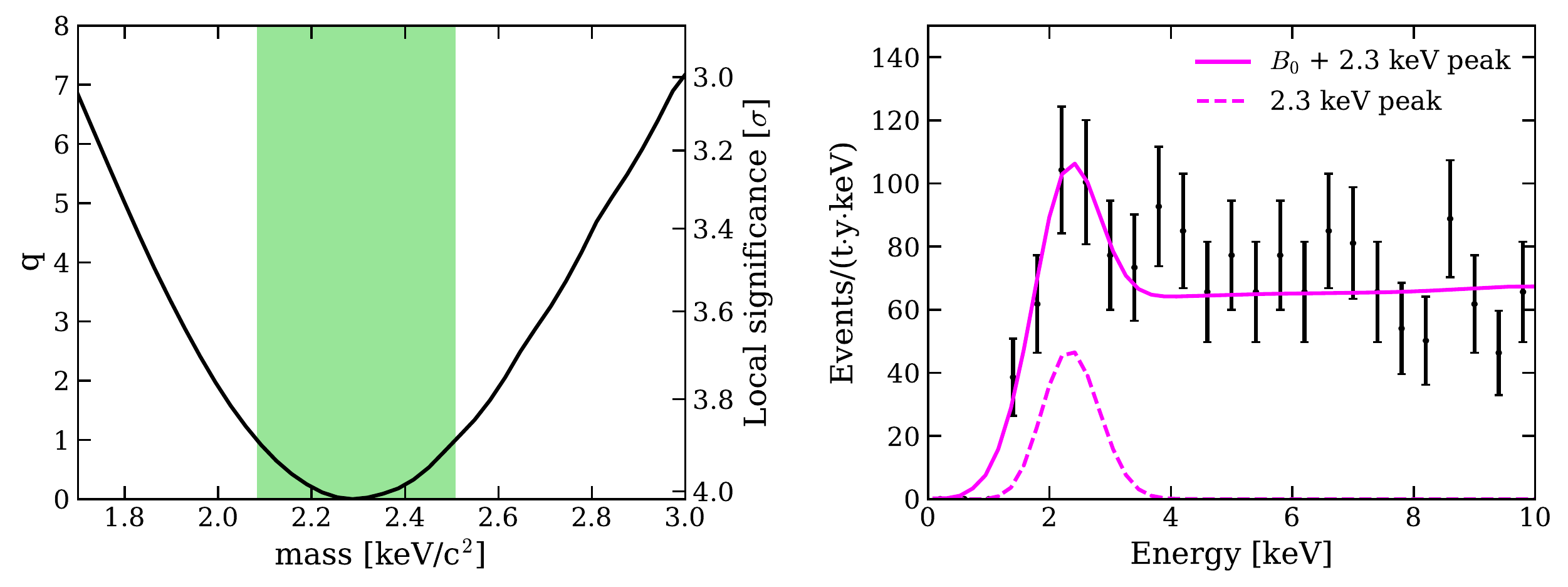}
    \caption{Left: The log-likelihood ratio $q$ for different bosonic dark matter masses with respect to the best-fit mass at 2.3\,keV/c$^2$. At each mass, we show the result for the corresponding best-fit coupling. The green band shows an asymptotic 68\% C.L.~confidence interval on the bosonic dark matter mass. The local significance for each mass is indicated by the right y-axis. Right: Best-fit of a 2.3\,keV peak and $B_0$ to the data. A 0.4\,keV binning is used for better visualization.}
    \label{fig:dm_peak}
\end{figure*}

These upper limits (90\% C.L.) are shown in Fig.~\ref{fig:dp_alps_limit}, along with the sensitivity band in green ($1\sigma$) and yellow\,($2\sigma$). The losses of sensitivity at 41.5\,keV and 164\,keV are due to the $^{83\mathrm{m}}$Kr and $^{131\mathrm{m}}$Xe backgrounds, respectively, and the gains in sensitivity at around 5 and 35\,keV are due to increases in the photoelectric \mbox{cross-section} in xenon. The fluctuations in our limit are due to the photoelectric \mbox{cross-section}, the logarithmic scaling, and the fact that the energy spectra differ significantly across the range of masses. For most masses considered, XENON1T sets the most stringent direct-detection limits to date on pseudoscalar and vector bosonic dark matter couplings.

Due to the presence of the excess, we performed an additional fit using the bosonic dark matter signal model, with the particle mass allowed to vary freely between 1.7--3.3 keV/c$^{2}$. The result gives a favored mass value of \mbox{(2.3 $\pm$ 0.2)\,keV/c$^2$} (68\% C.L.) with a 3.0\,$\sigma$ global (4.0\,$\sigma$ local) significance over background. A log-likelihood ratio curve as a function of mass is shown in Fig.~\ref{fig:dm_peak}~(left), along with the asymptotic 1-$\sigma$ uncertainty and the local significance for each mass. The spectral fit of the 2.3\,keV peak is illustrated in Fig.~\ref{fig:dm_peak}~(right). Since the energy reconstruction in this region is validated using $^{37}$Ar calibration data, whose distribution has a mean value within $<1\%$ of the expectation at 2.82\,keV~\cite{ar37_energy}, this analysis can also be used to compare the data to potential mono-energetic backgrounds in this region.

\subsection{Additional Checks}\label{subsec:sr2}
Here we describe a number of additional checks to investigate the low-energy excess in the context of the tritium, solar axion, and neutrino magnetic moment hypotheses.

The time dependence of events with energies in \mbox{(1, 7)\,keV} in SR1 was investigated and found to be inconclusive. The event rate is slightly higher in the beginning of SR1, but the rate evolution is statistically consistent with (1) a constant rate, (2) a constant background rate ($B_0$) plus a subtle $\sim7$\,\% (peak-to-peak) rate modulation from the change in Earth-Sun distance, and (3) a constant background rate~($B_0$) plus an exponentially decreasing component with a fixed half-life of 35\,days ($^{37}$Ar half-life) or 12.3\,years ($^3$H half-life). As another test of time dependence, we split SR1 into three periods with equal exposure and fit the data in each period with the ABC solar axion signal model. Similarly to the \mbox{(1,7)\, keV} rate evolution, the best-fit signal rate is the highest in the first period of SR1, but is not statistically significant as the signal rate is consistent within uncertainty between the three periods. We therefore conclude that, due to limited statistics, at this time we cannot use time dependence to exclude any of the hypotheses discussed in this work. More detailed time dependence studies will be presented in a forthcoming publication.

Since the excess events have energies near our 1\,keV threshold, where the efficiency is $\sim$ 10\%, we considered higher analysis thresholds to check the impact of this choice on the results. With the excess most prominent between 2 and 3\,keV, where the respective detection efficiencies are $\sim$ 80\% and 94\%, changing the analysis threshold has little impact unless set high enough so as to remove the events in question. This is not well-motivated, given the high efficiency in the region of the excess. For all thresholds considered (namely, 1.0, 1.6, 2.0, 3.0 keV), the solar axion model gives the best fit to the data. We hence conclude that our choice of analysis threshold impacts neither the presence nor interpretation of the low-energy excess.

We also checked data from Science Run 2 (SR2), an R\&D science run that followed SR1, in an attempt to understand the observed excess. Many purification upgrades were implemented during SR2, including the replacement of the xenon circulation pumps with units that (1) are more powerful, leading to improved purification speed, and (2) have lower $^{222}$Rn emanation, leading to a reduced $^{214}$Pb background rate in the TPC~\cite{magpump,Rn_emanation_inprep}, which is further decreased by online radon distillation. The resulting increased purification speed and reduced background make SR2 useful to study the tritium hypothesis. If the excess were from tritium (or another non-noble contaminant), we would expect its rate to decrease due to the improved purification; on the other hand, the rate of the signal hypotheses would not change with purification speed.  

While the SR2 purification upgrades allowed for an improved xenon purity and a reduced background level, the unavoidable interruption of recirculation for the upgrades also led to less stable detector conditions. Thus, in addition to a similar event selection process as SR1 in Sec.~\ref{subsec:dataselection}, we removed several periods of SR2 for this analysis to ensure data quality. Periods where the electron lifetime changed rapidly due to tests of the purification system were removed to reduce uncertainty in the energy reconstruction. We also removed datasets during which a $^{83\mathrm{m}}$Kr source was left open for calibration. Data within 50\,days of the end of neutron calibrations were also removed to reduce neutron-activated backgrounds and better constrain the background at low energies. After the other selections, this data would have only added $\sim$ 10\,days of live time; thus, for simplicity, it was removed rather than fit separately like the SR1 dataset. With these selections, the effective SR2 live time for this analysis is 24.4\,days, with an average ER background reduction of 20$\%$ in \mbox{(1, 30)\,keV} as compared to SR1. 

A profile likelihood analysis was then performed on SR2 with a similar background model as SR1, denoted as $B_{\mathrm{SR2}}$.  Since we are primarily interested in using this data set to test the tritium hypothesis, we focus on the tritium results.

\begin{figure}[!htbp]
    \centering
    \includegraphics[width=1\linewidth]{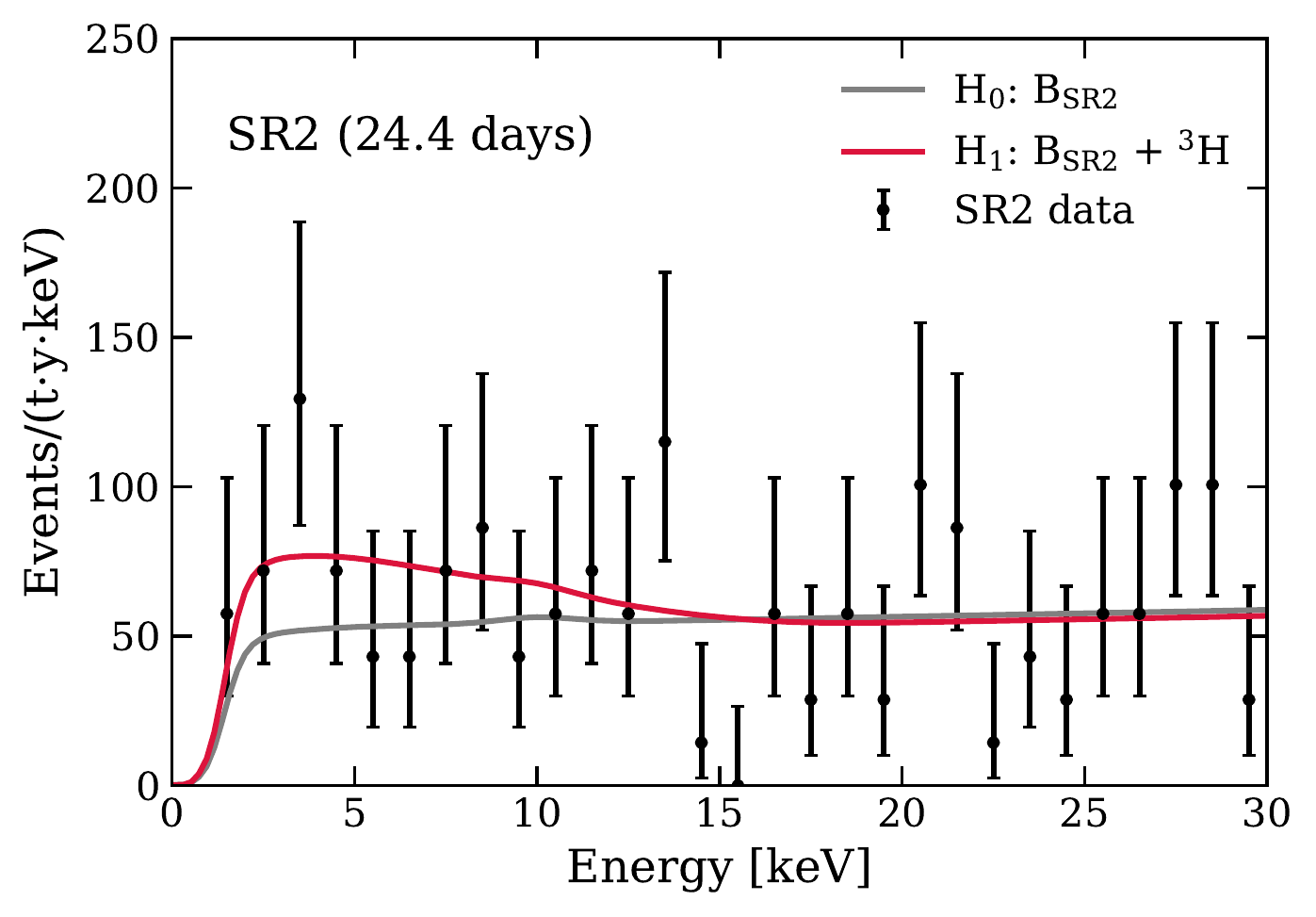}
    \caption{A fit to SR2 data if tritium is treated as a signal. The red (gray) line is the fit with (without) tritium in the background model.}
    \label{fig:sr2_fit}
\end{figure}

\begin{figure}[!htbp]
    \centering
    \includegraphics[width=1\linewidth]{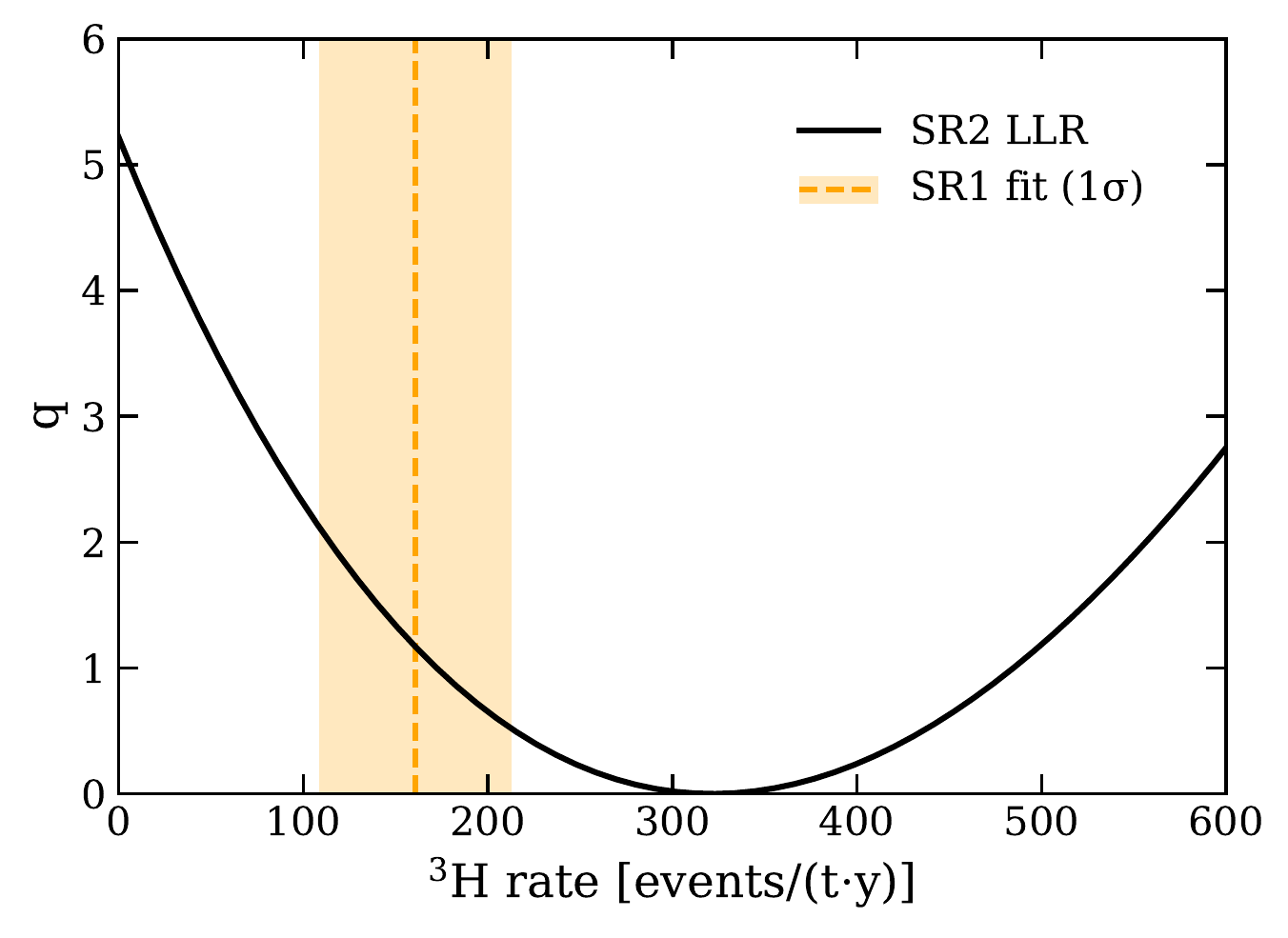}
    \caption{The log likelihood ratio curve for the tritium rate in SR2. The orange line and band indicate the best-fit and 1\,$\sigma$ uncertainty for the tritium rate in SR1. The SR2 fit result is consistent with SR1, but with a large uncertainty due to limited statistics.}
    \label{fig:sr2_llr}
\end{figure}

Similarly to SR1, we search for a tritium signal on top of the background model $B_{\mathrm{SR2}}$, and find that the background-only hypothesis is slightly disfavored at 2.3\,$\sigma$. The SR2 spectrum, along with the fits for the null ($B_{\mathrm{SR2}}$) and alternative ($B_{\mathrm{SR2}} + {^3\mathrm{H}}$) hypotheses, can be found in Fig.~\ref{fig:sr2_fit}. A log-likelihood ratio curve for the tritium component is given in Fig.~\ref{fig:sr2_llr}, which shows that the fitted tritium rate is \mbox{320 $\pm$ 160\,events/(t$\cdot$y)}, higher than that from SR1 but consistent within uncertainties. The rate uncertainty in SR2 is much larger than that in SR1 due to limited statistics. The solar axion and magnetic moment hypotheses give similar results, with significances $\sim 2 \,\sigma$ and best-fit values larger than, but consistent with, the respective SR1 fit results. Thus these SR2 studies are largely inconclusive. 

Lastly, we also checked these hypotheses in a different energy region using the so-called `S2-only' approach, where the requirement for an S1 signal is dropped, allowing for a $\sim 200$\,eV energy threshold. XENON1T's S2-only analysis~\cite{1t_s2only} was used to place limits on the tritium rate ($<$ 2256 events/(t$\cdot$y)) and $g_{\mathrm{ae}}$ ($<$ 4.8$\times10^{-12}$) that are far greater than, and therefore consistent with, the constraints derived here. The S2-only analysis is not as sensitive to the tritium and axion signals because both spectra peak above 1\,keV. On the other hand, many of the predicted signal events from neutrino magnetic moment fall below 1\,keV as the rate increases with falling energy, so the S2-only search is more relevant for this hypothesis. It yields a 90\% C.L. one-sided limit of $\mu_{\nu} < 3.1\times 10^{-11}\,\mu_B$, consistent with the upper boundary of the 90\% confidence interval obtained in Sec.~\ref{subsec:nmmresult}. Therefore, none of the discussed hypotheses are in conflict with the S2-only result.

\section{Discussion}
\label{sec:discussion}

We observe an excess of electronic recoil events at low energies in XENON1T data. In the reference region of \mbox{1--7\,keV}, 285 events are observed whereas \mbox{$232 \pm 15$\,events} are expected from the background-only fit to the data. The $\upbeta$ decay of tritium is considered as a possible explanation, as it has a similar spectrum to that observed and is expected to be present in the detector at some level. We are unable to independently confirm the presence of tritium at the $O(10^{-25})$\,mol/mol concentration required to account for the excess, and so treat it separately from our validated background model. If electronic recoils from tritium decay were the source of the excess, this would be its first indication as an atmospheric source of background in LXe TPCs. The tritium hypothesis clearly represents a possible SM explanation for the excess, but\,---\,based on spectral shape alone\,---\,the solar axion model is the most favored signal by the data at 3.4\,$\sigma$, albeit at only $\sim$\,2\,$\sigma$ if one considers tritium as an additional background.

If this excess were a hint of a solar axion, our result would suggest either (1) a non-zero rate of ABC axions or (2) a non-zero rate of both Primakoff and $^{57}$Fe axions. If we interpret the excess as an ABC axion signal (i.e., take $g_\mathrm{a\upgamma}$ and $g_\mathrm{an}^\mathrm{eff}$ to be zero), the required value of $g_{\mathrm{ae}}$ is smaller than that ruled out by other direct searches but has a clear discrepancy with constraints from indirect searches~\cite{white_dwarf_review, redgiants_gae}. These constraints are a factor of \mbox{$\sim$ 5--10} lower than reported here, although subject to systematic uncertainties. It is noteworthy that some of these astrophysical analyses, while their constraints are still stronger than direct searches, do in fact suggest an additional source of cooling compatible with axions~\cite{white_dwarf_review,axion_hunters}. If the indirect hints and the XENON1T excess were indeed explained by axions, the tension in $g_{\mathrm{ae}}$ could be relieved by underestimated systematic uncertainties in, e.g., stellar evolution theory~\cite{redgiants_gae} or white dwarf luminosity functions~\cite{Bertolami_2014}, or by a larger solar axion flux than that given in~\cite{Redondo:2013wwa}.

Although not considered in this work, XENON1T is also directly sensitive to the axion-photon coupling via the inverse Primakoff effect, whereby a solar axion coherently scatters off the effective electric field of the xenon atom, thus producing an outgoing photon and inducing an electronic recoil. This detection channel was first considered only recently for xenon-based detectors in~\cite{Dent} and~\cite{Gao}, which demonstrated that the tension of axion-photon coupling between the XENON1T excess and stellar constraints can be significantly reduced.  

Continuing to interpret the excess as a hypothetical QCD axion signal, we can extend the analysis to make statements on the axion mass $m_\mathrm{a}$ under assumptions of different models, as outlined in Sec.~\ref{subsec:solaraxion}. As examples, we consider a DFSZ model with variable $\beta_{\mathrm{DFSZ}}$ and KSVZ model with variable electromagnetic anomaly $E$ (for simplicity we fix the color anomaly $N=3$). Comparing these two classes of models with our 90\% confidence surface, we find that both are consistent with our result for a subset of parameters. For the DFSZ model, we find \mbox{$m_\mathrm{a} \sim 0.1-4.1~\mathrm{eV}/\mathrm{c}^2$} and \mbox{$\cos^2\beta_{\mathrm{DFSZ}} \sim 0.01-1$} would be consistent with this work. Alternatively, under the KSVZ model \mbox{$m_\mathrm{a} \sim 46-56~\mathrm{eV}/\mathrm{c}^2$} and $E=6$ would be similarly consistent. These model-specific mass ranges are not confidence intervals, as their specific assumptions were not included when constructing Fig.~\ref{fig:solax_limit}. We instead report a single, model-independent confidence region on the couplings to allow comparison with a variety of models, not just the examples mentioned here. 

Additionally, we describe a direct search for an enhanced neutrino magnetic moment. This signal also has a similar spectrum to the excess observed, but at 3.2\,$\sigma$ displays a lower significance than that from solar axions. We report a confidence interval of \mbox{$\mu_{\nu} \in (1.4,~2.9) \times 10^{-11}\,\mu_B$}\,(90\% C.L.), the upper boundary of which is very close to the world-leading direct limit reported by Borexino~\cite{borexino_nmm}. This shows that dark matter experiments are also sensitive to beyond-SM physics in the neutrino sector. Here we only search for an enhanced neutrino-electron cross-section due to an anomalous magnetic moment, but a similar enhancement would also occur in neutrino-nucleus scattering~\cite{Harnik_2012}. With the discrimination capabilities of LXe TPCs to ER and NR events, it would be interesting to consider this channel in future searches as well.

\begin{figure}[!htbp]
    \centering
    \includegraphics[width=1\linewidth]{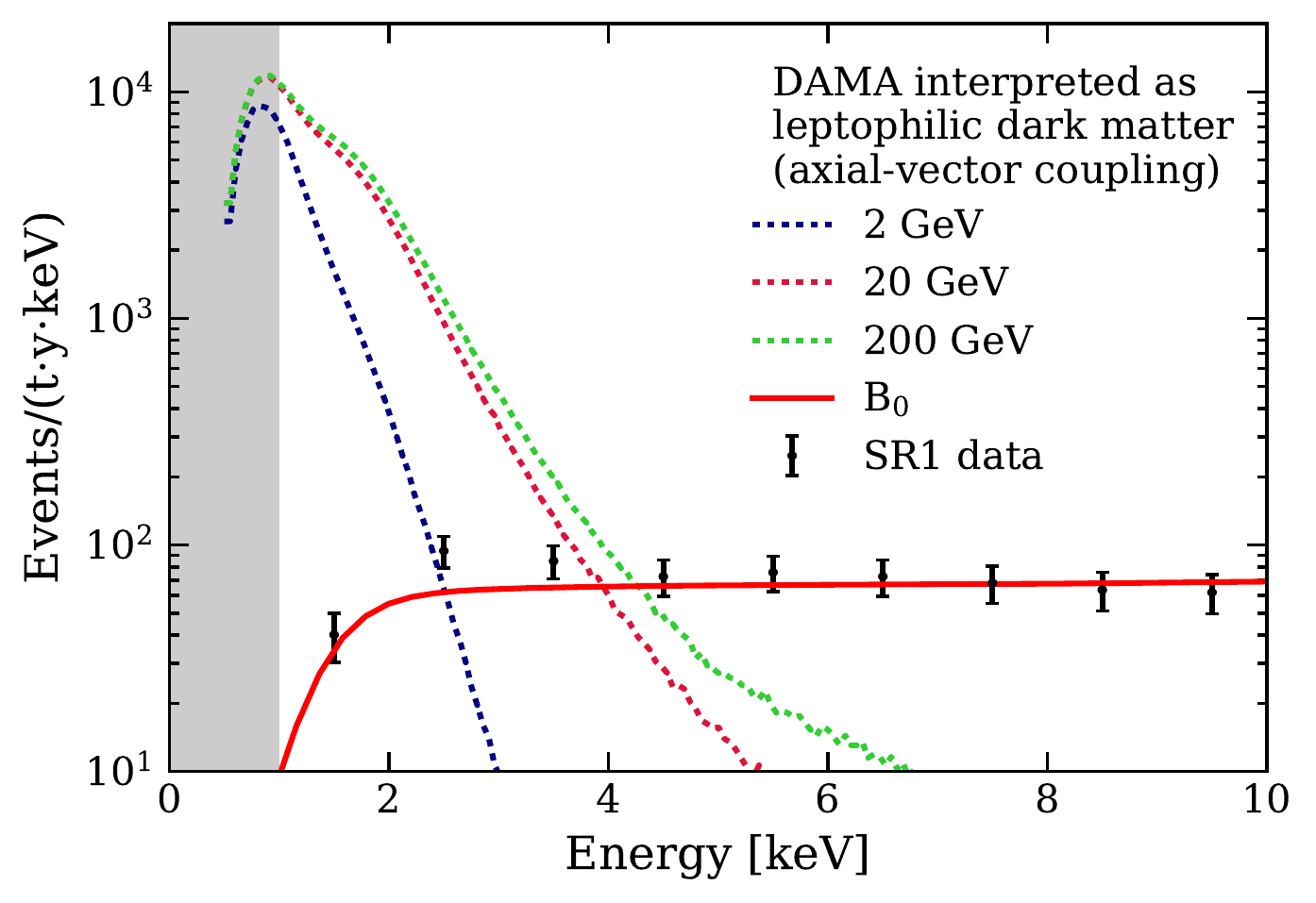}
    \caption{Comparison between DAMA expected signals and XENON1T data (signal plus background). Dotted lines represent the expected signal spectra of selected masses in the XENON1T detector if the DAMA modulated signals are interpreted as WIMPs scattering on electrons through axial-vector interactions. XENON1T data are indicated by black points and the background model~$B_0$ is illustrated by the red line. The right bound of the shaded region shows the threshold in this analysis.}
    \label{fig:axial_vector}
\end{figure}

If from an astrophysical source, the excess presented here is different from the result reported by the DAMA experiment, which claims that an observed annual modulation of events between 1 and 6~keV might be due to a dark matter signal~\cite{dama2008,Bernabei:2018yyw}. We present here a leptophilic dark matter model, where WIMPs couple with electrons through an axial-vector interaction~\cite{axialvector2009}. This model was used to explain the DAMA signal but was rejected already by the XENON100 experiment~\cite{xenonscience2015}. Interpreting the modulating source observed by DAMA under this model, the expected signal rate in the XENON1T detector would be more than 2 orders of magnitude higher than the total event rate we observed, as shown in Fig.~\ref{fig:axial_vector}. Consequently, the excess observed in this work is unrelated to the one observed by DAMA.

\section{Summary}
\label{sec:summary}

We report on searches for new physics using \mbox{low-energy} electronic recoils in XENON1T. In a search for bosonic dark matter, world-leading constraints are placed on the interaction strengths of pseudoscalar and vector particles. An excess is observed at low energies that is consistent with a solar axion signal, a bosonic dark matter signal with a mass of 2.3 keV/c$^2$, a solar neutrino signal with enhanced magnetic moment, or a possible tritium background. We are unable to confirm nor exclude the presence of tritium at this time. 

In an attempt to understand the low-energy excess, we performed a number of additional studies. The analysis of an additional data set called SR2\,---\,which displays a \mbox{$\sim$20\%} lower background rate but only \mbox{$\sim$10\%} statistics compared to SR1\,---\,is consistent with the SR1 analysis but largely inconclusive about the nature of the excess. An S2-only search, which is able to probe sub-keV energies, similarly yielded consistent constraints for all the discussed hypotheses. Compared to the excess observed by DAMA, it is much lower in rate and thereby unrelated.

The signals discussed here can be further explored in the next-generation detectors, such as the upcoming PandaX-4T~\cite{pandax4t}, LZ~\cite{lz2020} and XENONnT~\cite{ntmc} experiments. The next phase of the XENON program, XENONnT, featuring a target mass of 5.9\,tonnes and a factor of $\sim$6 reduction in ER background, will enable us to study the excess in much more detail if it persists. Preliminary studies based on the best-fit results of this work suggest that a solar axion signal could be differentiated from a tritium background at the 5\,$\sigma$ level after only a few months of data from XENONnT.

\vspace{20pt}

\textbf{Acknowledgements.} We thank Dr.~Roland Purtschert at Climate and Environmental Physics, University of Bern, for measurements of the $^{37}$Ar concentration at LNGS and for useful information regarding its production and transport. We also thank Dr.~Robin Gr{\"o}ssle of the Institute for Nuclear Physics - Tritium Laboratory at Karlsruhe Institute of Technology (KIT) and Dr. Richard Saldanha of Pacific Northwest National Laboratory (PNNL) for informative discussions. We gratefully acknowledge support from the National Science Foundation, Swiss National Science Foundation, German Ministry for Education and Research, Max Planck Gesellschaft, Deutsche Forschungsgemeinschaft, Netherlands Organisation for Scientific Research (NWO), Weizmann Institute of Science, ISF, Fundacao para a Ciencia e a Tecnologia, R\'{e}gion Pays de la Loire, Knut and Alice Wallenberg Foundation, Kavli Foundation, JSPS Kakenhi in Japan, the Abeloe Graduate Fellowship and Istituto Nazionale di Fisica Nucleare. This project has received funding or support from the European Union’s Horizon 2020 research and innovation programme under the Marie Sklodowska-Curie Grant Agreements No. 690575 and No. 674896, respectively. Data processing is performed using infrastructures from the Open Science Grid,  the European Grid Initiative and the Dutch national e-infrastructure with the support of SURF Cooperative. We are grateful to Laboratori Nazionali del Gran Sasso for hosting and supporting the XENON project.


\appendix
\section{$\upbeta$ Spectra Modeling}\label{sec:appendix}

This appendix briefly describes the different theoretical models used in the present work to compute the $\upbeta$ spectra for $^{214}$Pb, $^{212}$Pb, and $^{85}$Kr. 

\subsection{\label{app:geant4}GEANT4 Radioactive Decay Module}

The Radioactive Decay Module (RDM) in GEANT4\footnote{Here we refer specifically to the current version 10.6; however the corrections described in this work have been implemented since at least version 9.5~\cite{Agostinelli:2002hh}.} simulates the decay of a given radionuclide using the nuclear data taken from an Evaluated Nuclear Structure Data File (ENSDF) \cite{tuli96}. The required $\upbeta$ spectra are generated in a dedicated class using an analytical model. The $\upbeta$ spectral shape, i.e. the unnormalized emission probability per electron energy, is derived from Fermi's golden rule as:
\begin{equation}
\label{appeq:spec}
\dfrac{\text{d}N}{\text{d}W} \propto pWq^2 F(Z,W) C(W) S(Z,W),
\end{equation}
with $Z$ the atomic number of the daughter nucleus. Here, $W$ is the total energy of the $\upbeta$ particle and is related to its kinetic energy $E$ by $W = 1 + E/m_e$, with $m_e$ the electron rest mass. The maximum energy $W_0$ is defined identically from the energy of the transition $E_0$. The $\upbeta$ particle momentum is $p = \sqrt{W^2 - 1}$ and the (anti)neutrino momentum is $q = W_0 - W$, assuming a massless particle\,($m_{\nu} = 0$). 

The Fermi function $F(Z,W)$ corrects for the static Coulomb effect of the nucleus on the $\upbeta$ particle. Considering the Coulomb field generated by a point-like nucleus, the Dirac equation can be solved analytically and the well-known expression of the Fermi function can be derived. GEANT4 follows the approximate expression of the Fermi function from \cite{wilk70}. 

The shape factor $C(W)$ takes into account the nuclear and lepton matrix elements. Assuming constant values of the lepton wave functions within the nuclear volume, one can demonstrate that allowed and forbidden unique transitions can be calculated without involving the structure of the nucleus. For an allowed transition, the shape factor is constant: $C(W)=1$. In GEANT4, first, second and third forbidden unique transitions are calculated following the approximate expressions given in \cite{kono66} that were established by considering the analytical solutions of the Dirac equation, the same as for the Fermi function. In any other case, the decay is treated as allowed.

The atomic screening effect corresponds to the influence of the electron cloud surrounding the daughter nucleus on the $\upbeta$ particle wave function. GEANT4 takes this into account following the most widespread approach set out by Rose in \cite{1936RO06} almost a century ago. For a $\upbeta$ electron, this effect is evaluated by subtracting from the particle energy $W$ a constant Thomas-Fermi potential $V_{0}$ which only depends on $Z$. This corrected energy $W' = W - V_{0}$ replaces $W$ in all the quantities required for the calculation of the spectral shape, except in the (anti)neutrino energy $q$ because this neutral particle is not affected by the Coulomb field. The parameterization of the potential used in GEANT4 is close to the prescription given in \cite{wilk70sc}. The screening correction is then given by:
\begin{equation}
S(W,Z) = \dfrac{p'W'}{pW} \times \dfrac{F(Z,W')}{F(Z,W)}.
\end{equation}
It is noteworthy that this correction can only be applied for $W \geq V_{0}$, which creates a non-physical discontinuity in the spectrum at $W = V_{0}$, as seen in Fig.~\ref{fig:Pb214_app}.

\subsection{\label{app:NDS}IAEA LiveChart}

The $\upbeta$ spectra available on the IAEA LiveChart website \cite{NDS} are produced with the first version of the BetaShape program \cite{Mou16}. The required information for each transition is taken from the most recent ENSDF file with results from the latest nuclear data evaluation~\cite{1992NNDC}. 

The physics model in BetaShape has already been detailed in \cite{2015MO10}, except for the atomic screening effect. The $\upbeta$ spectral shape is described in the Behrens and B{\"u}hring formalism \cite{Behrens82} by: 
\begin{equation}
\label{appeq:spec}
\dfrac{\text{d}N}{\text{d}W} \propto pWq^2 F(Z,W) C(W) S(Z,W) R(Z,W),
\end{equation}
with all quantities as defined before. The quantity $R(Z,W)$ are the radiative corrections described below. 

In this formalism, the Fermi function is defined from the Coulomb amplitudes $\alpha_{k}$ of the relativistic electron wave functions:
\begin{equation}
\label{appeq:fermi}
F(Z,W) = F_0 L_0 = \dfrac{ \alpha^2_{-1} + \alpha^2_{+1} }{ 2p^2 }.
\end{equation}
These wave functions are numerical solutions of the Dirac equation for the Coulomb potential of a nucleus modeled as a uniformly charged sphere. Indeed, no analytical solution exists even for such a simple potential; however, the method from \cite{Behrens82} allows for a precise, and fast, calculation of the Coulomb amplitudes. The method inherently accounts for the finite nucleus size while other methods usually require an analytical correction ($L_0$ in Eq.~(\ref{appeq:fermi})).

The total angular momentum change $\Delta J = |J_i - J_f|$ and the parity change $\pi_i \pi_f$ between the initial and final nuclear states are from the input ENSDF file and determine the nature of the transition. Given that $L = 1$ if $\Delta J = 0$ or $1$ for an allowed transition, and $L = \Delta J$ for any $(L - 1)^{\text{th}}$ forbidden unique transition, the theoretical shape factor can be expressed as:
\begin{equation}
\label{appeq:shapfac}
C(W) = (2L-1)!~\sum_{k=1}^{L}{ \lambda_{k} \dfrac{p^{2(k-1)}~q^{2(L-k)}}{(2k-1)![2(L-k)+1]!} }.
\end{equation}
The $\lambda_{k}$ parameters are defined from the Coulomb amplitudes $\alpha_{k}$ by:
\begin{equation}
\label{appeq:lambda}
\lambda_{k} = \dfrac{ \alpha_{-k}^2 + \alpha_{+k}^2 }{ \alpha_{-1}^2 + \alpha_{+1}^2 }.
\end{equation}

In the case of forbidden non-unique transitions, the structures of the initial and final nuclear states must be taken into account, which greatly complicates the calculation. The usual approximation consists of treating such a transition as a forbidden unique transition of identical~$\Delta J$. The validity of this approximation, minutely tested in \cite{2015MO10}, can be demonstrated only for some first forbidden non-unique transitions, which are then calculated as allowed. Its generalization to every forbidden non-unique transition is implemented in BetaShape.

The radiative corrections are non-static Coulomb corrections from quantum electrodynamics. They can be split into two parts: the inner corrections, which are independent of the nucleus; and the outer corrections, which depend on the nucleus. Only the latter depend on the $\upbeta$\,particle energy. The outer radiative corrections $R(Z,W)$ take into account the internal bremsstrahlung process, by which the $\upbeta$ particles lose energy in the electromagnetic field of the nucleus. For allowed transitions, analytical corrections were derived in \cite{1967SI,1972JA} and are implemented in the first version of BetaShape as described in \cite{2015MO10}.

Finally, the spectral shape is modified by applying the screening correction $S(W,Z)$. The BetaShape program includes an analytical correction based on the work of B{\"u}hring \cite{Buhring1984} that is more precise than Rose's correction. The most realistic, spatially varying screened potentials at the time were of Hulth{\'e}n type (see \cite{Beh69} and references therein). B{\"u}hring first developed a version of the Dirac equation that correctly includes Hulth{\'e}n's potentials but simplified the angular momentum dependency, allowing analytical solutions to be established \cite{Buh83}. He then performed in \cite{Buhring1984} a radial expansion at the origin of both the wave functions and the Coulomb potential, including Hulth{\'e}n's screened potential, and retained only the dominant term. This procedure allows the determination of screened-to-unscreened ratios for the Fermi function $F_{0}L_{0}$ and the $\lambda_{k}$ parameters, which are then used to correct for screening in Eq.~(\ref{appeq:spec}). Therefore, the quantity $S(Z,W)$ in Eq.~(\ref{appeq:spec}) is more a symbolic notation. In BetaShape, this approach is used with Salvat's screened potentials \cite{Salvat87}, which can be expanded at the origin as:
\begin{equation}
\label{appeq:salvatexp}
V(r) = -\frac{\alpha Z}{r} + \frac{\alpha Z}{2} \beta + O(r),
\end{equation}
where $\beta$ is determined from the parameters $A_{i}$ and $\alpha_{i}$ given in \cite{Salvat87}:
\begin{equation}
\label{appeq:salvatbeta}
\beta = \sum\limits_{i=1}^{3}{A_{i} \alpha_{i}}.
\end{equation}
These potentials are widely used for their precision and completeness. It is noteworthy that B{\"u}hring's correction does not create any non-physical discontinuity in the spectrum as in Rose's correction. However, it tends to greatly decrease the emission probability at low energy.

\subsection{\label{app:improv}Improved Calculations}

When high precision at low energy is required, the modeling of $\upbeta^-$ decays must include the atomic screening and exchange effects. The two approximate screening corrections previously described are not sufficient. The exchange effect is even more significant and comes from the indistinguishability of the electrons. The regular, direct decay corresponds to the creation of the $\upbeta$ electron in a continuum orbital of the daughter atom. In the exchange process, the $\upbeta$ electron is created in an atomic orbital of the daughter atom and the atomic electron which was present in the same orbital in the parent atom is ejected to the continuum. This process leads to the same final state as the direct decay, i.e. one electron in the continuum, and is possible because the nuclear charge changes in the decay.

Precise relativistic electron wave functions are necessary to calculate such effects. The numerical procedure was described in detail in \cite{Mou14}, with the nucleus modeled as a uniformly charged sphere. For the continuum states, the Coulomb potential includes the appropriate Salvat screened potential. The wave functions, and therefore the Fermi function $F_{0}L_{0}$ and the $\lambda_{k}$ parameters, inherently take into account the screening effect. For the bound states, an exchange potential has to be added to this Coulomb potential and a specific procedure was implemented to ensure good convergence to precise atomic energies. In \cite{Mou14}, the one-electron energies from \cite{Des73} were considered while in the present work, the more accurate orbital energies from \cite{Kot97} that include electron correlations are used. 

A precise description of the exchange effect was set out in detail in \cite{Pyper88,Harston92}, but only for the allowed transitions. In such a case, $\upbeta$ electrons are created in continuum states with quantum number $\kappa = \pm 1$ and the selection rules imply that exchange can only occur with atomic electrons of identical $\kappa$, i.e. in $s_{1/2}$ ($\kappa=-1$) and $p_{1/2}$ ($\kappa=+1$) orbitals. The influence of the exchange effect can then be taken into account through a correction factor on Eq.~(\ref{appeq:spec}):
\begin{equation}
\frac{\text{d}N}{\text{d}W} \longrightarrow \frac{\text{d}N}{\text{d}W} \times \left( 1 + \eta_{\text{ex}}^{T} \right).
\end{equation}
The total exchange correction is defined by:
\begin{equation}
\eta_{\text{ex}}^{T} (E) = f_{s} (2 T_{-1} + T_{-1}^2) + (1 - f_{s}) (2 T_{+1} + T_{+1}^2),
\end{equation}
with:
\begin{equation}
f_{s} = \dfrac{ g_{-1}^{c'}(R)^2 }{ g_{-1}^{c'}(R)^2 + f_{+1}^{c'}(R)^2 }.
\end{equation}
All primed quantities refer to the daughter atom, and to the parent atom otherwise. The large and small components of the relativistic electron wave functions, respectively $g_{\kappa}^{c}$ and $f_{\kappa}^{c}$ for the continuum states and $g_{n,\kappa}^{b}$ and $f_{n,\kappa}^{b}$ for the bound states, respectively, are calculated at the nuclear radius $R$. The quantities $T_{-1}$ and $T_{+1}$ depend on the overlaps between the bound states of the parent atom and the continuum states of the daughter atom with energy $E$,
\begin{equation}
T_{(\kappa=-1)} = - \sum_{(n,\kappa)'}{}{ \langle (E\kappa)'|(n \kappa) \rangle \dfrac{g_{n,\kappa}^{b'}(R)}{g_{\kappa}^{c'}(R)} }
\end{equation}
and
\begin{equation} 
T_{(\kappa=+1)} = - \sum_{(n,\kappa)'}{}{ \langle (E\kappa)'|(n \kappa) \rangle \dfrac{f_{n,\kappa}^{b'}(R)}{f_{\kappa}^{c'}(R)} }.
\end{equation}
The sums are running over all occupied orbitals of the daughter atom of same quantum number $\kappa$.

It is noteworthy that in \cite{Mou14}, only the $s_{1/2}$ orbitals were taken into account, following the prescription in \cite{Harston92}. The ``new screening correction'' proposed in \cite{Mou14} was necessary to reproduce the experimental $\upbeta$ spectra of $^{63}$Ni and $^{241}$Pu, but was later found to be incompatible with a rigorous derivation of the $\upbeta$ spectrum starting from the decay Hamiltonian and the corresponding $S$ matrix. If correct screening and exchange effect with $s_{1/2}$ and $p_{1/2}$ orbitals are considered, together with precise atomic orbital energies, excellent agreement over the entire energy range of the two spectra is obtained.

Finally, more precise radiative corrections have been considered compared with those previously described. They were developed using more recent mathematical techniques and a significant change in the correction terms was found \cite{Cza04}. Describing the various changes is out of the scope of the present work; however, many details can be found in \cite{Hay18}. The influence of these new radiative corrections on the integrated $\upbeta$ spectrum is given for twenty superallowed transitions in \cite{Tow08}, for which an excellent agreement is obtained with the present implementation. It appears that these corrections are significantly smaller than the previous ones, especially for high atomic numbers.

\subsection{\label{app:appli}Application to the Transitions of Interest}

These different models have been applied to the ground-state to ground-state transitions in $^{212}$Pb, $^{214}$Pb, and $^{85}$Kr decays. The resulting spectra are similar to each other in the major part of the energy range, except at low energy.

\begin{figure}[!htbp]
    \centering
    \includegraphics[width=1\linewidth]{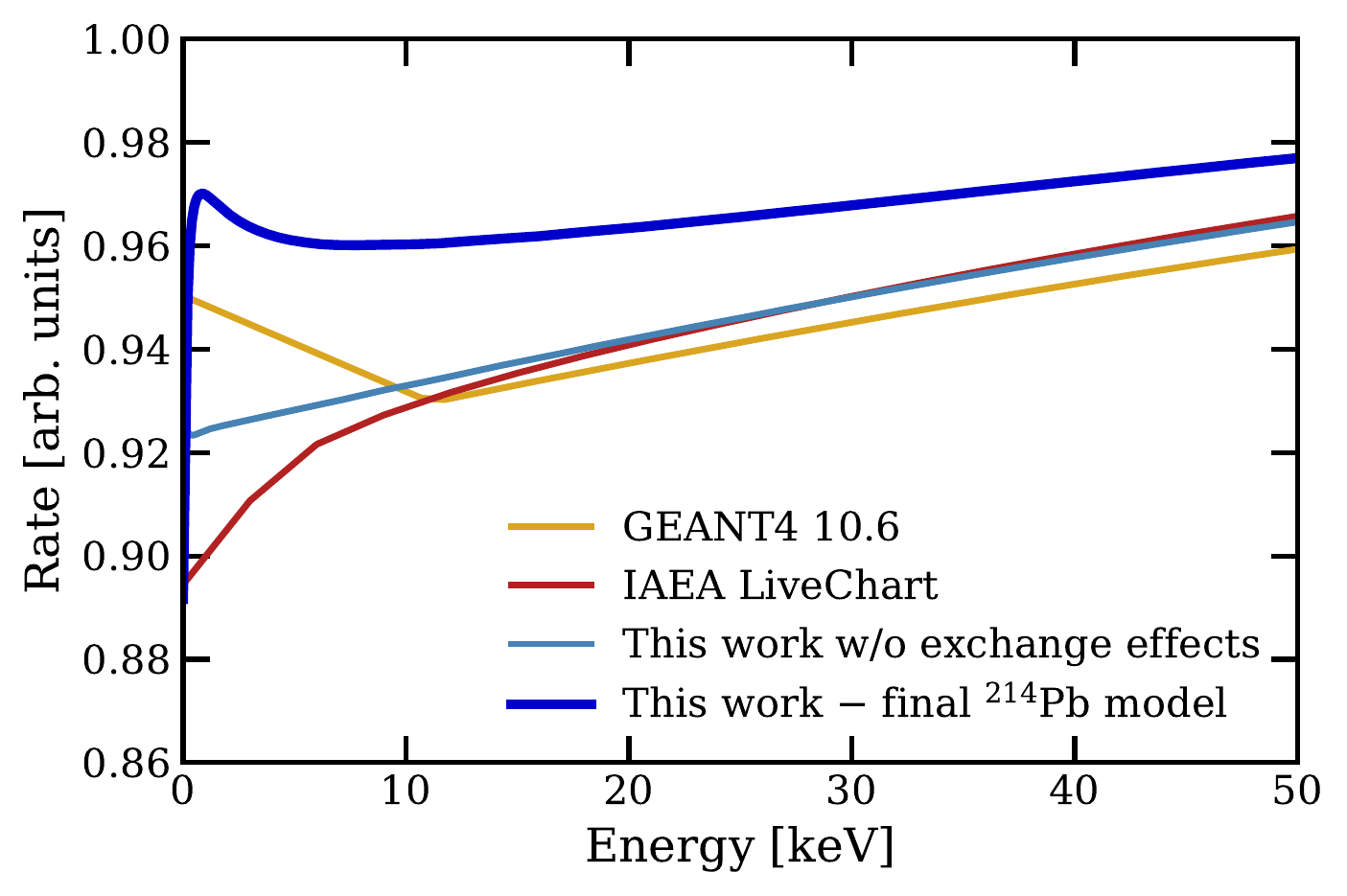}
    \caption{Low energy part of the $\upbeta$ spectral shape of the ground-state to ground-state transition in $^{214}$Pb decay. This first forbidden non-unique transition was calculated as allowed in every case but with different levels of approximations, as described in the text. The four spectra are normalized by area over the full energy range. See text for details on
    the shape of each spectrum.} 
    \label{fig:Pb214_app}
\end{figure}

The differences are illustrated in Fig.~\ref{fig:Pb214_app} for the low energy region of the $^{214}$Pb $\upbeta$ spectrum. The yellow curve is the GEANT4 RDM model as described in \ref{app:geant4} and the non-physical discontinuity due to the screening correction is clearly visible at 12\,keV. The red curve is from IAEA LiveChart, thus generated by the first version of the BetaShape program as described in \ref{app:NDS}. One can see the effect of B{\"u}hring screening correction that tends to decrease the emission probability. The cyan and blue curves were determined as described in \ref{app:improv}, without and with the atomic exchange correction, respectively. The screening effect is found to have a much smaller influence on the spectral shape when determined using a full numerical procedure than when applying an analytical approximation. However, the atomic exchange effect has a strong influence, as expected from previous studies \cite{Mou14}.

Both transitions in $^{212}$Pb and $^{214}$Pb ground-state to ground-state decays were calculated as allowed, accordingly with the approximation described in \ref{app:NDS}. It is important to keep in mind that formally, such first forbidden non-unique transitions should be determined including the structure of the initial and final nuclear states, a much more complicated calculation that is beyond the present scope. 

The transition in $^{85}$Kr decay is first forbidden unique and can thus be calculated accurately without nuclear structure. The description of the exchange effect from \cite{Pyper88,Harston92} used here is only valid for allowed transitions. For a first forbidden unique transition, one can expect a contribution of the $\kappa = \pm 2$ atomic orbitals but the exact solutions have still to be derived. However, the spectral shape is derived from a multipole expansion of the nuclear and lepton currents, as shown in the shape factor in Eq.~(\ref{appeq:shapfac}). Therefore, one can expect that the allowed exchange correction should give the main contribution, and this was done to determine the $^{85}$Kr $\upbeta$ spectrum.

The tritium $\upbeta$ spectrum used in this work was obtained from the IAEA LiveChart~\cite{NDS}, thus calculated using the standard Fermi function without corrections. As $^{3}$H decays via an allowed transition, this spectrum is sufficiently precise at energies above 0.5 keV, as confirmed experimentally in~\cite{Simpson}.

\subsection{\label{app:unc}Uncertainties}

The dominant contribution to the continuous XENON1T low-energy background comes from $^{214}$Pb $\upbeta$ decay. We thus focus the uncertainty discussion on the $^{214}$Pb ground-state to ground-state transition, calculated for the final model (blue curve in Fig. \ref{fig:Pb214_app}). 

The transition energy is directly given by the Q-value~\cite{Wan17}: $Q_{\beta} =$ 1018(11)\,keV. This uncertainty can be propagated by calculating the spectrum at \mbox{(1018 $\pm$ 11) keV}, namely at $1\sigma$. The result is an envelope centered on the spectrum calculated at the Q-value, which provides an uncertainty on the emission probability for each energy bin. The relative uncertainty is 1.7\% below 10\,keV and 1.1\% at 210\,keV. However, most of this uncertainty is removed because the $^{214}$Pb spectrum is left unconstrained in the fitting procedure. The remaining uncertainty component on the emission probability is $\sim 0.5$\% for each energy bin, in which the shape of the spectrum cannot vary steeply.

The atomic screening effect only slightly modifies the shape of the $\upbeta$ spectrum. Its uncertainty contribution can thus be safely ignored. The atomic exchange effect strongly affects the spectral shape below 5\,keV, and its accuracy depends on the atomic model used. For the $\upbeta$ spectra of $^{63}$Ni and $^{241}$Pu, the residuals between their high-precision measurement and the improved calculation in \ref{app:improv} showed that the agreement is better than the statistical fluctuations due to the number of counts in each energy channel, from 0.5\,keV to the endpoint energy. A conservative value of 1\% for each energy bin is the maximum relative uncertainty and is the value adopted here.

The $^{214}$Pb transition of interest is first forbidden non-unique. As explained in \ref{app:appli}, the nuclear structure should be taken into account for such a transition because it has an influence on the spectral shape. Treating it as an allowed transition induces an inaccuracy which cannot be estimated by comparison with a measured spectrum -- no measurement has been reported so far. In the same mass region, the $^{210}$Bi decay exhibits also a first forbidden non-unique, ground-state to ground-state transition with a comparable Q-value, and an experimental shape factor is available. As can be seen in \cite{Mou16b}, treating this transition as allowed leads to an important discrepancy with measurement. The question is then how this observation can be used for assessing an uncertainty to the $^{214}$Pb spectral shape. 

First, allowing the rate normalization to be free in the (1, 210) keV region absorbs the vast majority of any difference. Second, the nuclear structures of $^{210}$Bi and $^{214}$Pb are not identical. The $^{210}$Bi decay can be seen as two nucleons in the valence space above the doubly-magic $^{208}$Pb core, with the initial configuration $(p,1h_{9/2})(n,2g_{9/2})$ and two protons in the $1h_{9/2}$ orbital in the final state. However, this picture is too simple to be accurate because the core is not really inert. Nucleons from the core can give contributions to the $\upbeta$ decay matrix elements, mainly through meson exchange effects and core polarization effects \cite{eji78}. In $^{214}$Pb decay, a single proton in the $1h_{9/2}$ orbital is present in the final state and in the initial state, six neutrons are spread over the orbitals of the valence space but tend to couple to each other through pairing and dominantly occupy the $2g_{9/2}$ orbital. Contributions from the core nucleons can be expected to be relatively small compared to the main $(n,2g_{9/2}) \rightarrow (p,1h_{9/2})$ transition. In addition, even though it is difficult to predict if the nuclear structure component shifts the spectrum to lower energies, as for $^{210}$Bi, or to higher energies, a steep variation at low energy is not realistic. 

To conclude, we conservatively estimate a relative uncertainty on the spectral shape of 5\% due to the nuclear structure component and an additional 1\% for the energy dependency of the relative uncertainty on the maximum energy. Thus a 6\% total uncertainty on the spectral shape is estimated for the $^{214}$Pb $\upbeta$-decay model in this work.

\bibliography{bibliography.bib}

\end{document}